\documentclass{article}

\usepackage{arxiv}

\usepackage{xcolor}
\usepackage{graphicx}
\usepackage[thinlines]{easytable}
\usepackage{booktabs}
\usepackage{tabularx}
\usepackage{listings}

\usepackage[pdfusetitle,unicode=true,breaklinks=true]{hyperref}

\graphicspath{{figs/}}

\title{HACI: A Haptic--Audio Code Interface to Improve Educational
  Outcomes for Visually Impaired Introductory Programming Students}

\author{Pratham Srivastav Gandhi \\
Department of Computer Science, University of Chicago}

\date{May 2024}

\newcounter{chapter}
\renewcommand{\thechapter}{\arabic{chapter}}
\newcommand{\chapter}[1]{%
  \refstepcounter{chapter}%
  \section*{Chapter\ \thechapter.\ #1}%
  \addcontentsline{toc}{section}{Chapter\ \thechapter.\ #1}}

\begin{document}

\maketitle


\begin{abstract}
This thesis introduces the Haptic-Audio Code Interface (HACI), an educational tool designed to enhance programming education for visually impaired (VI) students by integrating haptic and audio feedback to compensate for the absence of visual cues. HACI consists of a non-resource-intensive web application supporting JavaScript program development, execution, and debugging, connected via a cable to an Arduino-powered glove with six integrated haptic motors to provide physical feedback to VI programmers. Motivated by the need to provide equitable educational opportunities in computer science, HACI aims to improve non-visual code navigation, comprehension, summarizing, editing, and debugging for students with visual impairments while minimizing cognitive load. This work details HACI's design principles, technical implementation, and a preliminary evaluation through a pilot study conducted with undergraduate Computer Science students. Findings indicate that HACI aids in the non-visual navigation and understanding of programming constructs, although challenges remain in refining feedback mechanisms to ensure consistency and reliability, as well as supplementing the current functionality with a more feature-reach and customizable accessible learning experience which will allow visually impaired students to fully utilize interleaved haptic and audio feedback. The study underscores the transformative potential of haptic and audio feedback in educational practices for the visually impaired, setting a foundation for future research and development in accessible programming education. This thesis contributes to the field of accessible technology by demonstrating how tactile and auditory feedback can be effectively integrated into educational tools, thereby broadening accessibility in STEM education.
\end{abstract}


\chapter{Introduction}
In today's digital age, programming skills are crucial for economic competitiveness and innovation. In fact, a study conducted by the Organization for Economic Co-operation and Development highlights that 90\% of future employment positions will require some level of digital literacy \cite{organisation2016skills}. This has been driven by the deep embedding of technology in our day-to-day lives, ranging from our smartphones and computers to home appliances, in-class education tools, vehicles, and more. Gen Z students (those born between 1995 and 2010) in particular, who are now learning introductory programming skills at the middle school, high school, and undergraduate levels \cite{schwieger2018reaching}, are true digital natives. That is, these students have used the internet, social media, video game platforms, mobile phones, and personal computers since their early ages \cite{bhalla2021digital}. Educators have noticed their Gen Z students, naturally curious about understanding the systems permeating their everyday lives, have expressed excitement about ``looking in" to the programs and digital devices around them \cite{abesadze2020make}. Beyond a natural curiosity to learn about such a ubiquitous concept, the non-partisan policy and governance think tank Brookings reports that educational systems worldwide have been integrating computer science (CS) education into their curricula in response to the high demand for CS professionals who enjoy stable, high-income careers \cite{vegas2020we}.

These career paths and educational opportunities are not equally available to all students, however. There is a significant inclusivity gap, particularly for visually impaired (VI) students. Specifically, research has shown that, in order to create ``globally competitive learners," students need to learn the ``4Cs" (creativity and innovation, critical thinking and problem solving, communication, and collaboration), as opposed to the ``3Rs" (reading, writing, and arithmetic) \cite{zain2016integration, alismail201521st}. In the realm of programming education, achieving the 4Cs learning objectives is done through project-based learning \cite{lamb2017key}. Educators have reported that enabling VI students to engage in programming project-based learning, however, currently faces the choke point of the lack of availability of tools designed with accessibility in mind \cite{pires2020exploring}. For example, Scratch, a visual block-based programming environment popular in introductory programming curricula, lacks compatibility with screen readers which VI learners frequently rely on \cite{fields2014programming, mountapmbeme2021teachers}. Visually impaired students often encounter significant barriers without accessible learning environments, ranging from simple syntactical comprehension to complex problem-solving scenarios where spatial and structural awareness of code is crucial \cite{alotaibi2020teaching}. 

While there have been efforts to make programming more accessible to VI learners, such as as the development of accessible block-based programming languages like Quorum \cite{stefik2017quorum} to address Scratch's accessibility shortcomings, these efforts predominantly focus on younger learners and simpler programming tasks. There is a notable scarcity of accessible tools designed for learning more advanced, text-based programming, which is crucial for higher education and professional development. While tools like StructJumper -- an Eclipse IDE plugin which creates a screen reader-friendly tree structure from Java code to allow easy non-visual navigation \cite{Baker2015} -- have been created for experienced VI programmers, more comprehensive tools also covering aspects of the programming learning process including debugging and code summarizing have not been created or widely implemented in classrooms.

This presents the opportunity to create the Haptic-Audio Code Interface (HACI), a tool designed to bridge this gap by providing a text-based programming environment that uses both audio and haptic feedback to make programming accessible and intuitive for visually impaired users by providing functionality to enhance non-visual code navigation, comprehension, summarizing, editing, and debugging. HACI leverages existing web technologies and simple, affordable hardware to create a programming environment specifically aimed at reducing the barriers faced by visually impaired programmers.

\section{Objectives}

The primary objectives of this thesis are to:
\begin{itemize}
    \item Develop a prototype of HACI, integrating haptic and audio feedback to support coding activities without the need for visual input.
    \item Evaluate the effectiveness of HACI in reducing the cognitive load and enhancing the understanding of code for visually impaired users.
    \item Explore the practical implications of implementing HACI in educational settings, including its acceptance among visually impaired students and its integration into existing curricula.
\end{itemize}

\section{Contributions}

This thesis makes several key contributions to the field of accessible education technologies:
\begin{enumerate}
    \item It presents a novel interface that combines haptic and audio feedback to aid visually impaired students in learning to program, which is one of the first of its kind tailored specifically for programming education.
    \item It provides empirical insights from a pilot study involving the use of HACI, offering preliminary evidence of its potential benefits and areas for improvement.
    \item It discusses the broader implications of multimodal learning tools in enhancing accessibility in education, suggesting pathways for future research and development.
\end{enumerate}

\section{Structure of the Thesis}

This thesis is organized as follows:
\begin{itemize}
    \item \textbf{Chapter 2} reviews the literature on challenges faced by visually impaired students in programming education and existing solutions.
    \item \textbf{Chapter 3} describes the design and technical implementation of HACI, detailing the integration of haptic and audio feedback mechanisms.
    \item \textbf{Chapter 4} outlines the methodology of the pilot study designed to evaluate HACI's effectiveness, including the adaptation for a case study with the author.
    \item \textbf{Chapter 5} presents the results from the pilot study, analyzing the potential impact of HACI on coding education for visually impaired students.
    \item \textbf{Chapter 6} discusses the findings, drawing conclusions about the efficacy of HACI and providing recommendations for future work.
    \item \textbf{Chapter 7} notes the limitations of the pilot study.
    \item \textbf{Chapter 8} proposes future improvements and additional features that can be incorporated into HACI and provides a study design that would allow a more robust assessment of HACI's effectiveness in addressing its goals.
    \item \textbf{Chapter 9} concludes the thesis, summarizing the research and its implications for future innovations in accessible programming education.
\end{itemize}

By addressing these objectives, this thesis aims to contribute to a more inclusive educational landscape, where visually impaired students can engage with computer science education on equal footing with their sighted peers, empowered by technology that caters to their specific learning needs.

\chapter{Background and Related Work}
\section{Challenges Faced while Using Text-Based Languages}
\label{sec:challenges}

While visually impaired (VI) developers can rely on screen readers to interact with development environments, a survey of experienced blind software developers found that many of the most commonly used programming environments, such as Eclipse, Microsoft Visual Studio, and XCode are either not compatible with screen readers or difficult to use through the standard audio feedback provided by a line-by-line screen reader \cite{Albusays2016}. This section presents the five primary challenges that previous work has found people with visual impairments encounter when working with text-based languages (TBLs). These challenges are Code Navigation, Code Skimming, Code Comprehension, Code Debugging, and Code Editing. While previous research has shown that these challenges affect both TBLs and block-based languages (BBLs), this thesis focuses on improving the accessibility of TBLs, and thus work on BBLs will be considered out of scope for this review.

\subsection{Code Navigation Challenges}
\label{sec:navchallenges}

Code Navigation challenges fundamentally drive many of the other challenges faced by VI programmers, as programmers must have a robust understanding of where they are working within a source code file, in the context of the other functions and code in the file, to effectively skim, comprehend, debug, and edit the program. Across multiple interviews, surveys, and observational studies, Code navigation was the most frequently reported challenge faced by VI programmers \cite{Albusays2016, Albusays2017, mealin2012exploratory}.

In interviews with 28 blind professional software developers focused on code navigation, Albusays et al. \cite{Albusays2017} found the line-by-line nature of screen readers to cause several Navigation issues. First, programmers reported issues looking for information elsewhere in a code base while maintaining their current position of focus. Sighted programmers can easily leave their cursor in a given position while scrolling to examine other parts of the source code visually. VI programmers, on the other hand, can only consume information from screen readers line-by-line, resulting in what VI programmers described as an inefficient and cognitively taxing process to find information in a code base \cite{huff2020examining}. Through interviews with eight VI programmers with varying degrees of blindness, ranging from no sight to basic shape and color identification, Mealin et al. \cite{mealin2012exploratory} also found this to be a commonly faced challenge, highlighting the presence of this obstacle across the VI programmer community.

This challenge can be exacerbated by the specific syntactic structure of the code, where single special characters may take up an entire line and interrupt the logical flow of the audible read-out of a program \cite{Albusays2016}. In the aforementioned set of 28 interviews, Albusays et al. \cite{Albusays2017} found that navigating nested structures, such as functions, loop/control statements, and if-statements was particularly difficult, as these are identified using alignment and special characters, which are either inaccessible or difficult to access through screen readers. Additionally, interviewees reported no feasible way to comprehend whitespace in cases where it is integral to the language's behavior, such as Python. Languages such as Ruby, which primarily employ text tokens and infrequently use non-alphanumeric characters, are preferable when only using a screen reader \cite{kane2014tracking}.

Once a programmer finds the information they are looking for, VI programmers have described encountering the additional task, which is difficult for the same reasons, of backtracking to the initial point of focus in the code \cite{Albusays2016}. This requires a symmetrically time-consuming process of searching for the initial focus position line-by-line. Alotaibi et al. \cite{Alotaibi2020} noted that this challenge frequently causes frustrations among VI developers, who often enter code at unintended positions. Mealin et al. \cite{mealin2012exploratory} found that VI programmers use temporary text buffers in separate windows of text editors or notepads to keep information on their code such as variable names, function signatures, and snippets of API documentation. This practice serves as an alternative to losing their focus in the file to navigate line-by-line to search for the same information. Other researchers also found the same \cite{huff2020examining}, adding that VI developers who use the text buffer approach employ screen readers and braille displays to enable this approach \cite{Albusays2017}. 

Navigation challenges extend beyond the source code itself. Integrated development environments (IDEs) often provide features to enhance programmer productivity. For example, Eclipse provides a package explorer for users to navigate through a code base represented as a tree structure, which  Smith et al. \cite{smith2003nonvisual} found to be inaccessible to most screen readers, and did not provide useful information to screen readers it was compatible with. The screen readers were only able to read out the current node selected in the tree and provided no information about how that tree was related to other nodes (sibling, child, etc.). VI programmers mentioned the inaccessibility of such IDE features in interviews conducted in \cite{Albusays2017, huff2020examining}. IDEs themselves also pose usability challenges due to poor accessible navigation. For example, Visual Studio doesn't generate any sounds or expose any functionality to a screen reader to make it clear when a developer switches between tabs \cite{stefik2011design}, while Eclipse does provide auditory feedback through a screen reader. Potluri et al. \cite{Potluri2018} noted the challenge of IDE feature discoverability which arises as a product of poor navigability in IDEs. For example, many features that are useful for specific contexts, such as variable watch windows (discussed in more depth in Chapter \ref{sec:debuggingchallenges}), are unknown to VI developers until others highlight their existence.

\subsection{Code Skimming Challenges}
\label{sec:skimchallenges}

Mountapembeme et al.'s meta-analysis of barriers for VI programmers \cite{Mountapmbeme2022} noted that Code Skimming challenges are understudied and often grouped with Code Navigation. For example, tools such as StructJumper \cite{Baker2015}, elaborated on in Chapter \ref{sec:audiosolutions}, have been created to help VI programmers efficiently navigate a large code base, but also help the programmer gain an overview by condensing the source code to an easily navigable structure that can be meaningfully understood through audio cues and screen readers. Code Skimming challenges are created by the same screen reader limitations which create Navigation challenges, as VI programmers cannot gain an overview of the code without going through it line-by-line \cite{mealin2012exploratory}. Without a purpose-built skimming tool, VI programmers usually consult API documentation to learn about the various functions through their function definitions and natural language descriptions. This serves as an effective alternative to the common IDE feature of code collapsing, which is inaccessible through a screen reader \cite{Albusays2017}.

\subsection{Code Comprehension Challenges}
\label{sec:compchallenges}

Code Comprehension challenges for VI programmers are similarly often studied in tandem with Code Navigation \cite{Mountapmbeme2022}. From the limited work done solely on Code Comprehension, it is clear that VI programmers face challenges created by relying on screen readers to comprehend code. Roberts and Weaver \cite{roberts2011audio} studied the use of both speech and non-speech audio in enabling VI programmers and found that VI programmers cannot effectively comprehend code using screen readers alone, and the use of non-speech cues greatly improves comprehension. Armaly et al. \cite{armaly2018comparison} studied the differences in how sighted and VI programmers read and summarize code. They found that VI programmers spend more time understanding function signatures than function bodies, and return to the signatures more often. Additionally, VI programmers focus less on function invocations themselves to comprehend code than their sighted counterparts. While Armaly et al. found that approaches to reading and summarizing code varied between sighted and VI programmers, they reported no perceived differences in the code summaries that were generated by their sighted and VI subjects. While this might seem contradictory to Roberts and Weaver's finding, which highlighted the ineffectiveness of a screen reader approach to Code Comprehension, it is important to note that Armaly et al.'s study used a maximum of 22 lines of code for subjects to summarize under no time restriction, focusing solely on the quality of the summaries. In reality, even code bases that introductory programming students face are larger, and the efficiency of comprehension is critical.

\subsection{Code Debugging Challenges}
\label{sec:debuggingchallenges}

Most debugging tools integrated into IDEs are highly visual and incompatible with screen readers \cite{Albusays2016, Albusays2017, stefik2009sodbeans}. VI programmers lack accessible options to understand the behavior of programs during execution. Additionally, many debugging cues available to sighted programmers during code editing, such as red squiggles to highlight syntax errors, yellow squiggles to highlight unused functions, and the conditional highlighting of terms based on their role, are inaccessible to VI programmers through screen readers\cite{Potluri2018}. Stefik et al. \cite{stefik2009sodbeans} found that Visual Studio 2005 exposed incorrect debugging information to screen readers while editing code due to a lack of accessibility considerations incorporated into its initial design. Additionally, during execution, the debugger provided no runtime information to screen readers. Potluri et al. further reported that runtime debugging tools such as Visual Studio's Watch Window, breakpoints, and expression and variable values are either incompatible with screen readers or difficult to access unless VI developers are specifically alerted to the actions needed to use them\cite{Potluri2018}. In interviews, Albusays et al. \cite{Albusays2016} confirmed both the difficulty in debugging before and during program execution. Multiple studies \cite{Albusays2016, Potluri2018} have noted that VI programmers rely on print statement-based debugging as a workaround to the runtime debugging issue, though it requires running their full code and observing run-time values of variables printed out in the console after execution has completed.

\subsection{Code Editing Challenges}

As mentioned earlier, due to Code Navigation challenges with preserving focus and backtracking, many VI programmers (Mealin et al. \cite{mealin2012exploratory} found this to be the majority of those interviewed in their study) use out-of-context editing instead of editing in-place. Interviewees in \cite{mealin2012exploratory} reported that editing code directly in the source code made them lose context. Thus, they would copy the code being edited to a text editor in a separate buffer window, edit it there, and paste it back into the source code at the location where their cursor had remained. Code Editing challenges can also be caused by ineffective screen readers even in programming environments that are themselves accessible to screen readers. This is particularly important to consider in educational contexts, where younger students may not be experienced in utilizing assistive technology, especially in complex contexts such as programming environments. For example, in a study surveying teachers of K-12 teachers of students with visual impairments \cite{mountapmbeme2020investigating}, teachers reported that code editing was one of the most difficult tasks for VI students in Swift Playgrounds due to the complexity of VoiceOver -- the native screen reader on iPad, the platform which runs Swift Playgrounds. While Swift Playgrounds is a hybrid of BBLs and TBLs, the students struggled specifically with text editing with VoiceOver.

\section{Audio-Based Solutions for VI Programmers}
\label{sec:audiosolutions}


One of the first programming tools designed for VI programmers is JavaSpeak \cite{smith2000java}. JavaSpeak is a purpose-built code editor designed for undergraduate introductory programming students with visual impairments to learn Java. Using keyboard shortcuts and audio cues, JavaSpeak provides an overview of source code and enables navigation. JavaSpeak's key contribution is its use of aural cues to provide the user with information about the code's syntax and organization at various granularity levels. These levels range from token-by-token (words and symbols) to compilation unit (i.e., `.java' source file). JavaSpeak generates the aural cues using IBM's ViaVoice text-to-speech software. Beyond aural cues, JavaSpeak's other key component is its navigation system, which uses the Windows JAWS key mappings for its functionality, allowing users to navigate through the program and select portions of the code to hear their aural representation.

The Auditory Code Overview and Navigation tool (ACONT) \cite{hutchinson2018initial} addresses code skimming and navigation challenges without pre-building an underlying structure of the source code, similar to JavaSpeak \cite{smith2000java}. Instead, ACONT uses ``timeline navigation," which allows users to peruse a class file line-by-line, with key bindings enabling swift jumps within the file; the 1, 2, and 3 keys jump to the start, middle, or end of a file while the arrow keys enable local navigation. ACONT employs a novel auditory representation of programming constructs through a blend of speech, non-speech sounds, and spearcons (compressed speech cues). For instance, an `if statement' might trigger a door-opening sound, embedding the navigation experience within an easily interpretable soundscape. In a comparative study assessing these audio cues, speech was deemed most accurate, but non-speech sounds were preferred for their simplicity and engaging nature, suggesting an optimal balance between informative and user-friendly auditory feedback in programming environments for the visually impaired. Ludi et al. also conducted a study \cite{ludi2016exploration} to determine which type of auditory cue improves VI programmers' ability to navigate and understand source code (although for a block-based language) and similarly concluded that speech performed the best, with spearcons performing equally well in some cases and earcons always performing the worst in both comprehension and navigation.

While JavaSpeak is an early standalone tool created for VI programmers, Aural Tree Navigator \cite{smith2003nonvisual} is one of the first plugins created to improve the usability of the Eclipse IDE for VI programmers, created by Smith et al. This tool generates a tree-like hierarchical structure of a program, navigable through keyboard shortcuts and aural speech cues similar to those used to operate JavaSpeak. The specific keyboard shortcuts and commands that Aural Tree Navigator uses were determined using extensive user studies. To address the Code Navigation challenges of tree structures mentioned in the discussion of Eclipse's package explorer in Chapter \ref{sec:navchallenges}, Aural Tree Navigator's speech output communicates all of the information that the Eclipse package explorer conveys visually, including a description of the node and its logical position within the tree.

In 2015, Baker et al. built on Aural Tree Navigator's work \cite{smith2003nonvisual} and created StructJumper \cite{Baker2015}, another Eclipse plugin that addresses three of the primary challenges discussed in Chapter \ref{sec:audiosolutions} Code Navigation, Code Skimming, and Code Comprehension. StructJumper creates an abstract syntax tree (AST) to represent a Java class as a hierarchical tree structure, where AST nodes represent classes, functions, control statements, etc. Where Aural Tree Navigator used custom keyboard commands for its audio output \cite{smith2003nonvisual}, StructJumper primarily focuses on solving the Code Navigation challenge while exposing relevant information to a third-party screen reader that a VI programmer may already be using. StructJumper instead uses keyboard commands solely for navigation, allowing the user to quickly move between the code and the AST node corresponding to that code, solving the focus preservation and backtracking challenges discussed in Chapter \ref{sec:navchallenges}. Navigating through the AST also allows users to skim a condensed representation of the source code and comprehend its functionality.


Building on the foundational concepts introduced by tools like StructJumper \cite{Baker2015}, AudioHighlight \cite{armaly2018audiohighlight} was created by Armaly et al. in 2018 as a specialized solution tailored to enhance code skimming for VI programmers, particularly within web-hosted environments. Distinct from StructJumper's broad focus and IDE integration, AudioHighlight focuses on the specific challenge of efficiently navigating and understanding code structures directly from the web. By employing HTML tags to demarcate key structural elements like classes, functions, and control statements, AudioHighlight creates an accessible, hierarchical representation of code that leverages the inherent compatibility of screen readers with HTML, thus facilitating a more intuitive navigation experience for visually impaired users. This approach deviates from the conventional AST generation, offering a web-native solution that integrates seamlessly with online code repositories. Armaly et al. empirically validated the efficacy of AudioHighlight through comparative studies against both GitHub's standard code presentation and StructJumper, enabled by an Eclipse IDE plugin version of AudioHighlight. Compared to using a screen reader with Github, the researchers showed AudioHighlight enabled VI programmers to complete tasks faster and more easily. AudioHighlight performed comparably to StructJumper, but participants reported being able to learn how to use AudioHighlight faster and complete tasks more slightly more quickly than when using StructJumper.




Potluri et al. developed CodeTalk \cite{Potluri2018} to address the critical aspect of code debugging. Distinguished from similar tools by its comprehensive approach, CodeTalk enhances code navigation and comprehension through a dual interface consisting of a tree structure and a summarized list of functions. This combination allows developers to swiftly transition between functions, offering a macroscopic view of the code's architecture. CodeTalk includes an audio debugger that introduces accessible alternatives to conventional debugging features. Through audio descriptions, developers can utilize ``talkpoints" akin to breakpoints in visual debuggers, enabling precise tracking and marking within the code. This feature is complemented by auditory cues that convey variable values and syntax errors, bypassing the need for visual indicators. Designed as a plugin for Visual Studio and supporting languages like C\# and Python, Potluri et al. validated CodeTalk through user studies, demonstrating that it significantly aids in debugging tasks.


The Wicked Audio Debugger (WAD) \cite{stefik2007wad} is another plugin focused on debugging, designed for Visual Studio 2005. WAD employs auditory feedback, utilizing plain speech to represent program constructs during the execution phase. The tool articulates key programming elements through speech, with particular attention to speech properties such as speed, tone, and the strategic use of pauses between utterances. These design considerations stem from comprehensive pilot studies conducted by Stefik et al. \cite{stefik2007wad}, the creators of WAD, aimed at exploring the optimal auditory conditions for debugging comprehension. The studies revealed that pauses between speech utterances significantly improved the understanding of the sonified program, leading to design enhancements in WAD that include the careful selection of vocabulary to reduce ambiguity and the introduction of nesting level information before the articulation of control flow statements. This methodical structuring allows programmers to grasp the hierarchical organization of code more effectively. The empirical outcomes from the pilot studies demonstrated that participants were able to accurately interpret approximately 86\% of dynamic program behavior through WAD's auditory feedback.


\section{Haptic and Tactile Solutions for VI Programmers}

Compared to audio-based solutions, much less work has been done on building solutions that VI programmers interact with physically -- haptic and tactile solutions -- to address the Code Navigation, Code Skimming, Code Comprehension, Code Debugging, and Code Editing challenges discussed in Chapter \ref{sec:challenges}. Despite the lack of many specific haptic tools created for VI programmers, previous work has shown that haptic interaction can effectively enable VI students to complete tasks that are traditionally highly visual. For example, Dorsey et al. developed a comprehensive approach to engaging visually impaired students with the tactile and auditory aspects of robotics through multi-modal interfaces integrating text-to-speech translators, audio feedback, and haptic interfaces  \cite{dorsey2014developing}. During several workshops, researchers utilized these tools to convert visually displayed information into audible formats, enabling students to grasp basic programming syntax and robot command libraries effectively. Furthermore, the introduction of haptic feedback through a Wii remote controller and auditory feedback through piano notes associated with different robot actions allowed students to perceive their robot's actions, such as distance traveled or encountering obstacles, thereby ``seeing" the outcomes of their programming efforts.

Capovilla et al. present a novel approach to imparting algorithmic concepts to visually impaired students through the use of toy-building bricks and plates \cite{capovilla2013teaching}. This method utilizes tactile models to represent data structures and algorithms, thereby offering a hands-on learning experience. The study conducted a field experiment with five blind participants, focusing on teaching basic search algorithms such as linear search, binary search, and lookup in a binary search tree through tactile engagement. The results demonstrated that participants could successfully grasp the algorithmic principles and apply them in practical tasks, indicating the effectiveness of tactile models in fostering a deeper understanding of algorithmic thinking. This approach not only aids in overcoming the visual barriers inherent in computer science education but also highlights the potential of tactile learning tools in enhancing cognitive comprehension of complex concepts for visually impaired learners, suggesting broader applicability in inclusive educational settings.


The most robust tactile tool for VI programmers is the Tactile Code Skimmer (TCS) \cite{falase2019tactile}, developed as a Visual Studio Plugin, which introduces a novel tactile approach to assist blind developers in overcoming code skimming challenges. TCS employs a physical device comprising six horizontal sliders that represent the indentation levels of code lines, thereby providing a tangible representation of the code's structure. This tactile method allows for the mapping of six lines of code at any given time, with code blocks at the same indentation level condensed onto a single slider, facilitating rapid navigation through a source file. Crucially, TCS addresses the focus preservation issue inherent in screen reader use, where visually impaired programmers often lose their place when navigating away from their current focus to understand code structure. By enabling users to maintain their cursor position in the Visual Studio Editor while exploring code indentation through tactile feedback, TCS significantly reduces the cognitive load and auditory information overload commonly associated with screen readers.








\chapter{HACI Design and Implementation}

\begin{figure}[h] 
    \centering
	\includegraphics[trim={0.0cm 0.0cm 0.0cm 0.0cm},clip,width=0.7\textwidth]{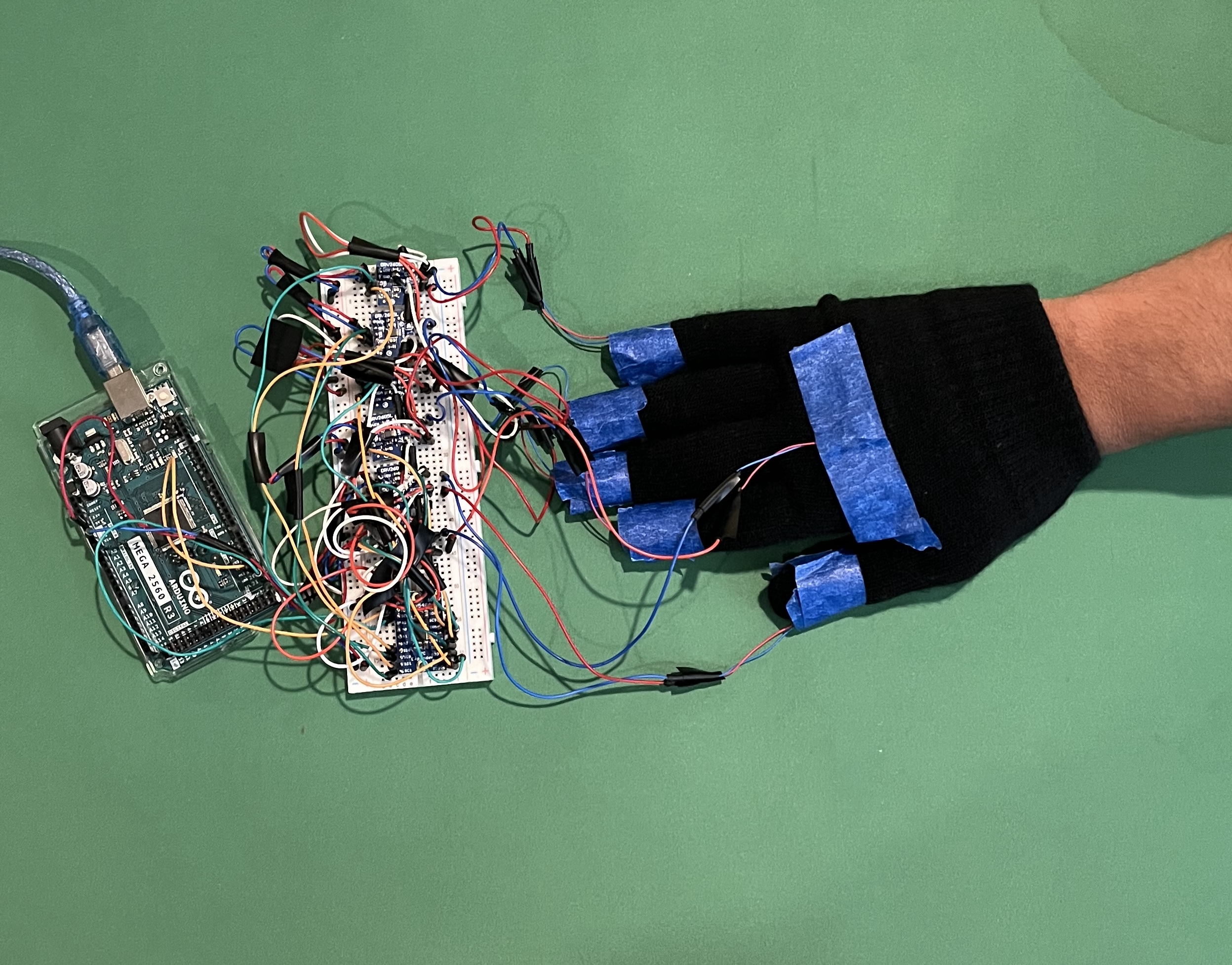}
	\vspace{-0.0em}
    \caption{Complete HACI hardware setup. Note the shorter wires connecting the glove motors to the hardware, which are for testing purposes and were elongated to allow comfortable and free hand movement during experimental use.\label{fig:hardwarepic}}
\end{figure}

To address the challenges discussed Chapter \ref{sec:challenges}, Haptic-Audio Code Interface (HACI) was created. HACI is a web-based JavaScript development environment. The choice of JavaScript as the HACI target language and the web environment for the HACI IDE in was driven by a couple of reasons. JavaScript is a widely accessible text-based language that can be run in any modern web browser, ensuring that teachers and students in educational settings with resource or technological (e.g., slow internet connection) constraints wouldn't be limited by the need to purchase and install costly and large tools. Additionally, building HACI as a web-based application ensured wider platform-agnostic support since the user interface is compiled and displayed by the browser itself and the code is compiled and run within the browser as well. Finally, by running in the browser, HACI supports future integration with other web-based technologies and easy connection to physical feedback devices through web sockets. All design decisions for HACI were made in support of 5 central design principles (DPs). These design principles and the decisions they drove are discussed below, with HACI input functionality summarized in Table \ref{tab:functionality} and HACI feedback summarized in Table \ref{tab:feedback}.

\begin{table*}[h]
    \centering
    \caption{HACI Input Functionality}
    \label{tab:functionality}
    \begin{tabularx}{\textwidth}{XX}
    \toprule
    \textbf{Command} & \textbf{Result} \\
    \midrule
    Ctrl-Shift-S & Toggle typing read-out on/off \\ \midrule
    Ctrl-B & Toggle granularity of typing read-out between characters and words \\ \midrule
    Ctrl-, & Drop cursor Marker 1 \\ \midrule
    Option-, & Jump to cursor Marker 1 \\ \midrule
    Ctrl+. & Drop cursor Marker 2 \\ \midrule
    Option-. & Jump to cursor Marker 2 \\ \midrule
    Option-1 & Jump to top/start of code \\ \midrule
    Option-2 & Jump to middle of code \\ \midrule
    Option-3 & Jump to bottom/end of code \\ \midrule
    Cmd-Enter & Executes code in Editor \\ \midrule
    Command + I & Switch to Errors Panel \\ \midrule
    Command + J & Switch to Code Panel \\ \midrule
    Command + K & Switch to Console/Terminal Panel \\ 
    \bottomrule
    \end{tabularx}
    
\end{table*}

\begin{table*}[h]
    \centering
    \caption{Feedback Provided to HACI Users}
    \label{tab:feedback}
    \begin{tabularx}{\textwidth}{XXX}
    \toprule
    \textbf{Triggering Command or Event} & \textbf{Sound Response} & \textbf{Haptic Response} \\
    \midrule
    Ctrl-G & Reads out current line &  \\ \midrule
    Ctrl-Number (e.g., Ctrl-1, Ctrl-2, etc.) & Reads out previous number of lines &  \\ \midrule
    Ctrl-V & Reads out name of function cursor is currently inside &  \\ \midrule
    Ctrl-E &  & Buzz indicates direction of error (middle finger buzz if error is above current line, center of hand buzz if error is below current line, center of hand double tap if error on current line)  \\ \midrule
    Navigate to indention increase (rightward) line &  & Ring finger buzzes \\ \midrule
    Navigate to indention decrease (leftward) line &  & Index finger buzzes \\ \midrule
    Navigating onto `[' or `\{' characters & Speaks ``open bracket'' or ``open brace'' & Thumb buzzes \\ \midrule
    Navigating onto `]' or `\}' characters & Speaks ``close bracket'' or ``close brace'' & Pinky finger buzzes \\ \midrule
    Navigating to a new line where an “if” statement begins & Door opening &  \\ \midrule
    Navigating to a new line where an “if” statement ends & Door slamming shut &  \\ \midrule
    Navigating to a new line where a loop begins & Car engine start &  \\ \midrule
    Navigating to a new line where a loop ends & Car brake screech &  \\ \midrule
    Navigating to a new line containing a syntax error & Speaking gibberish &  \\ \midrule
    Navigating to a new line containing a runtime error & Pitter-patter of running feet & \\
    \bottomrule
    \end{tabularx}
    
\end{table*}

\section{DP1: Minimize Extraneous Cognitive Load of the Programmer}
\label{sec:cogload}

The HACI user interface consists of three panels -- a main panel to edit code in, a smaller panel in the top half of the right column for errors, and a panel in the bottom half of the right column for the console. These components will be referred to as Code Editor, Error Panel, and Console, respectively. It is important to note that, for this study which focuses primarily on the audio and haptic feedback (not visual), the UI was not a primary concern. 

As discussed in Chapter \ref{sec:navchallenges}, VI programmers reported that the constraint of consuming information line-by-line required them to move their cursor to explore code and lose track of their initial focus position. To address this, HACI allows users to drop two distinct markers at the current location of their cursor, navigate away, and return to the position of either of the markers by using the corresponding keyboard shortcut. As additionally covered in Chapter \ref{sec:navchallenges}, visually impaired programmers often used a separate Notepad scratchpad to write code to maintain their cursor position in the codebase and noted the friction in their workflow created by navigating between multiple applications. To improve this, HACI presents the Code Editor, Error Panel, and Console on the same screen and allows simple navigation between the panels using keyboard shortcuts whose layout on the keyboard corresponds to the panels' layout in the UI.

\section{DP2: Effectively Convey the Structure of Code During Navigation}
\label{sec:codeStructDP}

Chapters \ref{sec:navchallenges}, \ref{sec:skimchallenges}, and \ref{sec:compchallenges} discussed several interrelated challenges visually impaired programmers face while trying to navigate, skim, and comprehend code using a typical screen reader. One of these challenges was the time cost programmers faced because of navigating their code line-by-line. HACI helps users save time by providing keyboard shortcuts to jump to the start, middle, or end of the current file, in addition to jumping to markers as discussed in Chapter \ref{sec:cogload}.

Programmers also reported difficulty in understanding the code as they navigated through it with a screen reader. This included friction caused by code structure meaning encoded by indentation which wasn't represented audibly. HACI assists users in understanding indentation by vibrating the user's right fourth finger when the user navigates to a line where the indentation from the previous line increases. Similarly, HACI vibrates the right index finger when navigating to a line with a decreased indentation from the line before it. Visually impaired programmers also reported the screen readers do not convey meaning encoded by non-alphanumeric characters that didn't have clean audio representations (such as ``\{" or ``//") well. HACI addresses this by including a specialized audio feedback map that encodes relevant JavaScript-specific characters and character combinations as readable phrases. For example, HACI reads ``//" as ``double slash comment," ``|" as ``pipe," and ``=$>$" as ``arrow function." A complete summary of this speech mapping is provided in Table \ref{tab:speechmapping}. Additionally, when the HACI user navigates onto a ``[", ``\{", ``]", or ``\}", and as one of these characters is read aloud in speech feedback, the HACI buzzes the user's right thumb if the character is an open bracket or brace and buzzes the user's right little finger.

\begin{table*}[h]
    \centering
    \caption{HACI Read-Out Symbol to Speech Mapping}
    \label{tab:speechmapping}
    \begin{tabularx}{\textwidth}{XX}
    \toprule
    \textbf{Symbol} & \textbf{Spoken Phrase} \\
    \midrule
    $\{$ & Open brace \\ \midrule
    $\}$ & Close brace \\ \midrule
    $($ & Open parenthesis \\ \midrule
    $)$ & Close parenthesis \\ \midrule
    $[$ & Open bracket \\ \midrule
    $]$ & Close bracket \\ \midrule
    $=$ & Equals \\ \midrule
    $==$ & Double equals \\ \midrule
    $===$ & Triple equals \\ \midrule
    $<$ & Less than \\ \midrule
    $>$ & Greater than \\ \midrule
    $!$ & Exclamation mark \\ \midrule
    $!=$ & Not equals \\ \midrule
    $\&\&$ & And \\ \midrule
    $||$ & Or \\ \midrule
    $/*$ & Start block comment \\ \midrule
    $*/$ & End block comment \\ \midrule
    $//$ & Double slash comment \\ \midrule
    $|$ & Pipe \\ \midrule
    $~$ & Tilde \\ \midrule
    $`$ & Backtick \\ \midrule
    $=>$ & Arrow function \\ \midrule
    $++$ & Increment \\ \midrule
    $--$ & Decrement \\ \midrule
    $<<$ & Left shift \\ \midrule
    $>>$ & Right shift \\ \midrule
    $!==$ & Strict not equals \\ \bottomrule
    \end{tabularx}
    
\end{table*}

\section{DP3: Provide Comprehensible Audible Snapshots of Current Position in Code}

To understand the context of the code they were currently working on, visually impaired programmers tediously navigated above and below their point of focus to have the code read to them line-by-line. To make this task simpler, HACI provides shortcuts to allow the user to have HACI read out the current line or read out a number of lines above the current line. HACI also adds tone/sound effect-based (as opposed to text-based) audio feedback to alert the user when they navigate to a new line where an ``if" statement or a loop begins or ends. That is, if statements are signified by the sound of a door opening and closing when the statement starts and ends, respectively, and loops are signified by the sound of a car engine starting up and a car brake screen when the loop begins and ends, respectively. Finally, based on the position of the cursor, HACI alerts the user ``You are in the function [function name]" and speaks the variable names consumed by the function or ``You are not inside a function,"  when the corresponding shortcut is pressed. Mealin et al. \cite{mealin2012exploratory} noted that, in interviews, visually impaired programmers specifically expressed interest in IDE functionality which allows them to attain information about the currently in-focus function while leaving their cursor in its current location. As discussed more broadly in Chapter \ref{sec:compchallenges}, VI programmers spend more time with function signatures than sighted programmers to comprehend code, which HACI supports through these features.

\section{DP4: Ensure Customizability and Provide Appropriate Feedback to Aid Code Writing}

Visually impaired programmers, and particularly the introductory-level students whom HACI was made for, have a variety of learning and code comprehension styles. To accommodate as many types of learners as possible, HACI provides various customization options. To help programmers avoid typing errors, HACI reads characters out loud as they are typed in the Code Editor. The granularity of this audio feedback can be toggled between tokens (words) and individual characters, and the granularity is applied across read-outs both while the user is typing and also when the user triggers a one-line or multi-line read-out. The while-typing read-outs can also be toggled on or off.

\section{DP5: Provide Non-Text Feedback to Aid Debugging Process}

The final guiding design principle for HACI was shaped by the difficulties visually impaired programmers reported in debugging existing code, as discussed in Chapter \ref{sec:debuggingchallenges}. To provide a solution for these challenges, HACI applies a combination of audio and haptic feedback to provide the user with an understanding of where the bug is and what type of error it creates. HACI has a keyboard shortcut which, while held, will trigger a haptic buzz to indicate the direction of the line of the first error that the browser ran into when last attempting to run the code in the Code Editor (this error is also shown in the Error Panel). A buzz on the user's right middle finger indicates that the error is above the cursor's current line, a buzz on the middle of the top of the user's right hand indicates the error is below the current line, and a double tap on the top of the user's right hand indicates the error is at the current line.

Once the user has navigated to the line containing the code causing the error, audio feedback will give them a fast initial impression of the type of error. That is, HACI will either play a sound bite of someone speaking ``gibberish" to indicate that there is a static (i.e., syntax) error on the current line or a sound bite of the pitter-patter of running feet to indicate that there is a runtime error on the current line. At this point, users may be able to debug the issue by simply having HACI read out the line, depending on the type of error. To get more details about a given error, however, users can use the appropriate shortcut to switch the UI focus on the Error Panel, and then hit a keyboard shortcut which will trigger an audio read-out of the detailed JavaScript error message.

\section{Technical Implementation}

The technical implementation of the Haptic-Audio Code Interface (HACI) is designed to offer an accessible programming environment by integrating intuitive user interface design with audio and haptic feedback mechanisms. The interface is developed using web technologies that facilitate interaction for visually impaired users. Key components include a user interface built with React.js, dynamic audio cues delivered through the Web Speech API and Howler.js for non-speech sounds, and responsive haptic feedback enabled by an Arduino-based hardware setup. These elements work in concert to provide a programming experience that is not only accessible but also enriches the user's understanding of the code structure and debugging process through multimodal feedback.

\subsection{User Interface and JavaScript Programming}

The HACI interface was constructed using React.js \cite{React}, a choice made for its component-based architecture, which facilitates the modular design of user interface elements like the code editor, console, and error panel. For the code editor functionality, the Ace Editor library was integrated due to its extensive API, which supports features crucial for accessibility, such as keyboard shortcuts and screen reader compatibility. Ace Editor was also chosen for its support for custom syntax highlighting, which was adapted to enhance code readability under the HACI paradigm.

To handle JavaScript code execution within the browser environment, HACI utilizes the Browserify tool to bundle Node.js-style modules for use in the browser, ensuring a seamless execution environment for user-written code. This is complemented by the use of Babel to transpile ES6+ JavaScript code to ensure backward compatibility with older browsers.

Keyboard navigation and shortcuts within HACI were implemented with the Mousetrap library, chosen for its lightweight nature and ease of binding keyboard commands to complex callback functions, allowing for an intuitive and responsive user interface optimized for visually impaired users.

To determine the context of the cursor within the code, particularly to identify the current function, HACI utilizes the \texttt{acorn} JavaScript library. Acorn is a small, fast, JavaScript-based parser capable of generating an Abstract Syntax Tree (AST) from ECMAScript code. The choice of Acorn was motivated by its compatibility with modern JavaScript syntax and its integration ease within a React.js environment. By generating an AST, HACI can accurately track the cursor's location relative to function blocks, allowing for context-aware audio and haptic and audio feedback.

\begin{figure}[h] 
    \centering
	\includegraphics[trim={0.0cm 0.0cm 0.0cm 0.0cm},clip,width=0.85\textwidth]{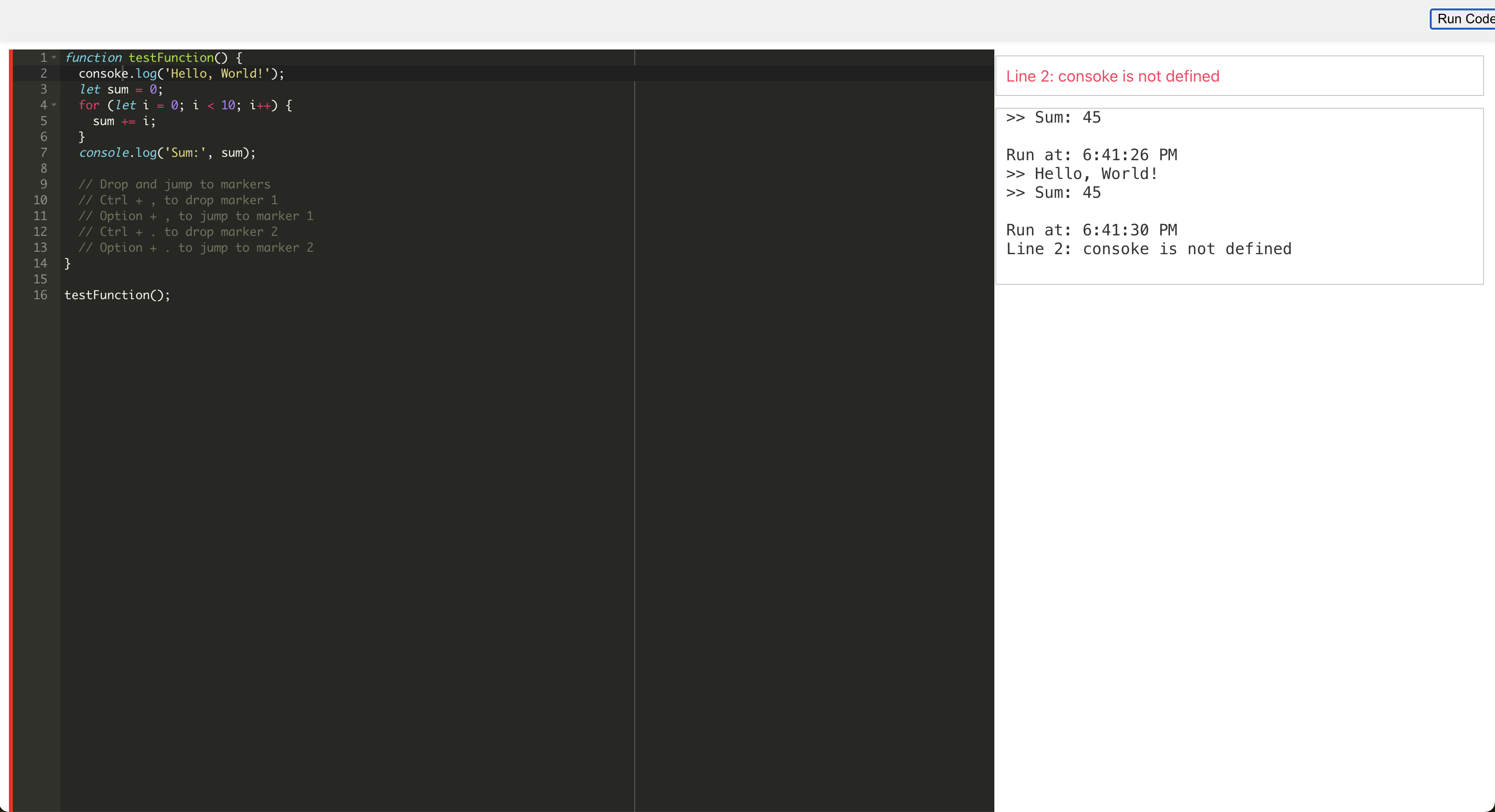}
	\vspace{-0.0em}
    \caption{Screenshot of HACI user interface with example code, a sample error, and sample console output.\label{fig:ui}}
\end{figure}

A screenshot of the user interface is provided in Figure \ref{fig:ui}

\subsection{Audio Feedback}

Audio feedback within HACI leverages the Web Speech API, specifically chosen for its native support in modern browsers and its comprehensive text-to-speech (TTS) capabilities. The API's flexibility allowed for the implementation of customized feedback settings, such as adjusting the verbosity of spoken feedback and introducing specific auditory cues for programming constructs.

For non-speech audio cues, such as environmental sounds indicating programming structures, the Howler.js library was employed. This decision was based on Howler.js's robust support for audio sprite manipulation, enabling efficient triggering of multiple sound effects without the overhead of managing numerous audio files or instances.

\subsection{Haptic Feedback}

The integration of haptic feedback in HACI is achieved through an Arduino Mega 2560 REV3 microcontroller, interfaced with the web application via the Web Serial API. This API was selected for its ability to establish a direct connection between web applications and serial devices, facilitating real-time communication essential for responsive haptic feedback. The Arduino sketch was developed to interpret signals from the web application and activate vibration motors accordingly.

\begin{figure}[h] 
    \centering
	\includegraphics[trim={0.0cm 0.0cm 0.0cm 0.0cm},clip,width=0.7\textwidth]{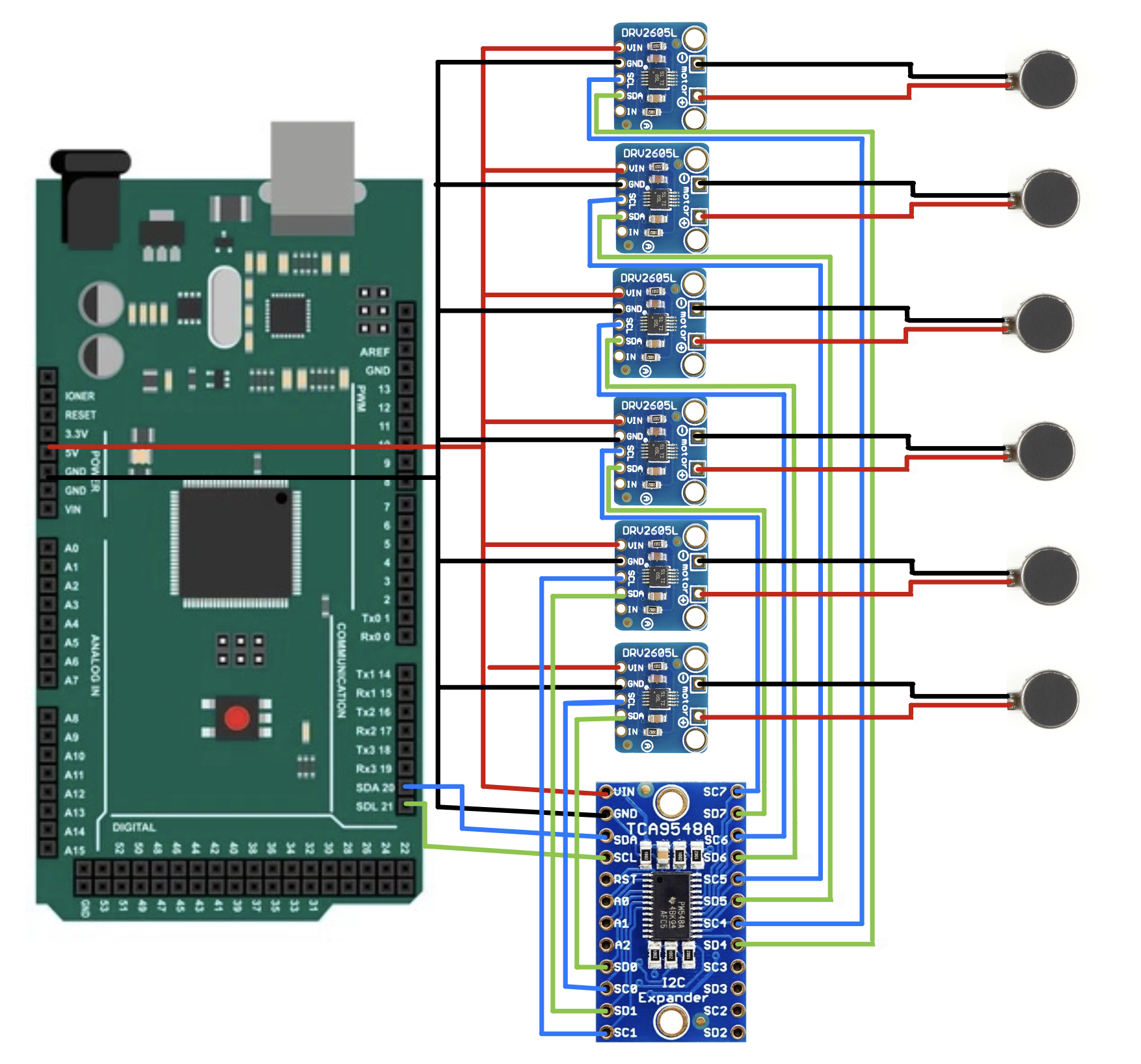}
	\vspace{-1.0em}
    \caption{Fritzing schematic of HACI hardware, including Arduino Mega, I$^2$C multiplexer, 6x DRV2605L haptic motor controllers, and 6 haptic motors.\label{fig:schematic}}
\end{figure}

The HACI haptic feedback glove includes 6 10mm by 3mm 3-Volt coin-shaped haptic motors mounted either on top of a finger or the top of the middle of the hand on a knit glove. The motors were controlled using DRV2605L motor drive modules, which received signals through I$^2$C data sent to the motor controllers from the Arduino's SDA and SCL pins. Since the Arduino only has one SDA and SCL pin each, and because the motor controller chips have fixed I$^2$C addresses, an Adafruit TCA9548A 1$^2$C Multiplexer was used to expand from just the I$^2$C address 0x70 to a dynamically adjustable range of 8 addresses between 0x70 and 0x77. The Multiplexer was then connected to the 6 motor controllers, which were in turn, wired to one haptic motor each. This is illustrated in the schematic provided in Figure \ref{fig:schematic}, and a photograph of the completed HACI hardware setup is provided in Figure \ref{fig:hardwarepic}.

To bridge the React application with the Arduino, the \texttt{react-serial} library was utilized, providing a React hook-based interface to interact with the Web Serial API, thus maintaining the React application's declarative nature while handling imperative serial communication.

\chapter{Pilot Study Design}
\label{chapter:pilotstudydesign}



\section{Participants}

This study enrolled 10 undergraduate students from the University of Chicago. Eligible participants were those who had completed at least one of the introductory Computer Science courses: CMSC 141, CMSC 151, or CMSC 161. This criterion ensured that participants possessed a foundational understanding of programming concepts, which was essential for the tasks they undertook during the study. The recruitment leveraged University-affiliated channels, including Registered Student Organizations (RSOs) and Computer Science Department forums, to reach potential participants. The demographics of recruited participants are provided in Table \ref{tab:demo}.
\\

\begin{table*}[h]
    \centering
    \caption{Demographics of Pilot Study Participants}
    \label{tab:demo}
    \begin{tabularx}{\textwidth}{XX}
    \toprule
    \textbf{} & \textbf{Count} \\
    \midrule
    Gender &  \\ 
    \hspace{3mm}Male & 6 \\ 
    \hspace{3mm}Female & 4 \\ \midrule
    Primary Major &  \\ 
    \hspace{3mm}Computer Science & 8 \\ 
    \hspace{3mm}Physics & 1 \\ 
    \hspace{3mm}Economics & 1 \\ \midrule
    Years of Study &  \\ 
    \hspace{3mm}1 year & 0 \\ 
    \hspace{3mm}2 years & 1 \\ 
    \hspace{3mm}3 year & 4 \\ 
    \hspace{3mm}4 years & 5 \\ \midrule
    Age &  \\ 
    \hspace{3mm}18-19 & 1 \\ 
    \hspace{3mm}20-21 & 6 \\ 
    \hspace{3mm}22-23 & 3 \\ 
    \hspace{3mm}24+ & 0 \\ \midrule
    Race &  \\ 
    \hspace{3mm}Asian & 5 \\ 
    \hspace{3mm}White & 3 \\ 
    \hspace{3mm}African American & 0 \\ 
    \hspace{3mm}Hispanic & 1 \\ 
    \hspace{3mm}Mixed Race & 1 \\ \midrule
    Total & 10 \\ \bottomrule
    \end{tabularx}
    
\end{table*}

\section{Set-Up}
The experimental sessions were conducted in a dedicated, sound-proof meeting room within the John Crerar Library on the University campus. This controlled environment was chosen to ensure privacy and minimize distractions, thus allowing participants to focus entirely on the tasks at hand while minimizing their risk of feeling judged. Each participant was equipped with the Haptic-Audio Code Interface (HACI), which includes the right-hand glove outfitted with small motors to provide haptic feedback. The experimental computer setup included all necessary software for interfacing with the HACI, alongside standard programming tools and environments necessary for JavaScript development, which were all built into the HACI UI.

\section{Procedure}
Upon arrival, each participant underwent a comprehensive briefing session, which included an overview of the study's goals and the informed consent process. For participants with limited or no prior experience with JavaScript, a succinct tutorial session was provided to introduce them to the critical syntax and concepts relevant to the subsequent tasks.

Following the introductory phase, participants were blindfolded to simulate a visual impairment condition and fitted with the HACI gloves. An initial training session was then conducted, guiding participants through a series of rudimentary code navigation, editing, and debugging tasks designed to familiarize them with the HACI's functionality.

The core of the experiment consisted of a series of structured programming tasks:
\begin{enumerate}
    \item \textbf{Code Summarizing:} Participants were presented with code segments where identifiers such as function and variable names were obscured. They were then asked to interpret the code's functionality and provide a verbal summary.
    \item \textbf{Function Implementation and Modification:} Participants created a new function from scratch and modified an existing `Main' function to incorporate the newly created function.
    \item \textbf{Debugging Tasks:} Participants tackled a set of debugging challenges involving multiple small functions, each containing a specific error, such as index out of bounds, misspelling of a variable, or use of an undeclared variable.
\end{enumerate}

These tasks were designed to evaluate the efficacy of the HACI in supporting code comprehension and problem-solving without visual cues. Each task was time-constrained, with participants encouraged to verbalize their thought processes. The complete interview guide which was used to conduct the pilot study interviews, including the guiding points for the interviewers, the specific interview questions, and the code content of each of the programming tasks, is presented in Appendix \ref{sec:interviewQuestions}. All sessions were audio and screen recorded, ensuring comprehensive data capture for subsequent analysis. The recordings were transcribed and anonymized to maintain participant confidentiality.

\section{Data Collection and Analysis Approach}

The study employed a mixed-methods approach to data collection, encompassing both qualitative and quantitative data to provide a comprehensive understanding of participants' experiences with the Haptic-Audio Code Interface (HACI). 

\subsection{Data Collection}
\hfill\\
Qualitative data was primarily collected through audio recordings of each session, capturing participants' verbalized thought processes, comments, and reactions while engaging with the programming tasks. Additionally, open-ended interviews conducted at the end of each session contributed further qualitative insights into participants' subjective experiences, challenges faced, and perceived utility of the HACI.

Quantitative data comprised of task completion times, error rates (i.e., how many times the compiler identified an error during task completion), and the number of interactions with specific HACI features (e.g., haptic feedback triggered, audio cues utilized, etc.). This data was automatically logged by the experimental software running the HACI, ensuring a precise and unobtrusive collection of participants' interactions with the interface.

\subsection{Data Analysis}
\hfill\\
Qualitative data analysis entailed transcribing the audio recordings from the experimental sessions and subsequent interviews. Thematic analysis was applied to these transcripts to identify recurring themes, patterns, and insights related to participants' experiences with the HACI. This involved a detailed coding process to categorize participants' feedback on usability, effectiveness of audio-haptic feedback, and overall engagement with the programming tasks.

Quantitative analysis focused on between-subject comparisons to understand how each participant's interaction with the HACI shaped their productivity. Metrics such as task completion times, error rates, and usage frequency of specific HACI features (e.g., particular haptic feedback or audio cues) were compiled for each task. These metrics were analyzed to identify trends and variations in participants' performance in the study potentially driven by the usage or non-usage of certain features. For example, faster task completion times or lower error rates for those who do use a HACI features vs. those who don't could indicate the effectiveness of that feature in enabling users to achieve their learning goals.

The integration of thematic insights from the qualitative analysis with the quantitative performance metrics provided a comprehensive understanding of the HACI's impact on facilitating programming tasks for participants. This approach ensured a nuanced interpretation of how the haptic-audio feedback system supports coding activities for visually impaired users.

\chapter{Results}
I now present results from the pilot study conducted with undergraduate Computer Science students, as described in Chapter \ref{chapter:pilotstudydesign}. Results are presented first by the aspect of HACI functionality they pertain to, followed by cross-cutting successes and issues highlighted during the study. The study provides insights into the potential and current limitations of HACI in supporting visually impaired programmers. While the system demonstrates a promising direction for accessible programming environments, the feedback mechanisms, particularly the haptic cues, highlighted the potential for further refinement to enhance reliability and precision. Additionally, participant feedback indicates scope for improvement in HACI's debugging functionality, particular in helping the programmer isolate the location of the error-causing term faster and easier. These improvements are essential for ensuring that HACI not only supports but enhances the programming experience for users with visual impairments.

\section{Cognitive Load}
\label{sec:cogLoadResults}

The use of HACI notably reduced the cognitive load for subjects by simplifying code navigation and syntax orientation. The shortcuts, such as \texttt{Option-1} for jumping to the top of the code, were particularly highlighted for enhancing efficiency. One participant mentioned, ``The quick jump shortcuts were lifesavers. I felt more in control and less stressed about keeping track of everything." Another participant similarly noted, ``Using control commands for navigation really streamlined my process. I didn't have to worry about losing my place as much." Beyond movement through different parts of the code, these shortcuts also helped minimize cognitive load by providing jumps to ``absolute" points such as the start, middle, and end of the code. This functionality ``eliminated any ambiguity about whether the cursor had reached the top yet after scrolling or not."

Participants also found the marker shortcuts (e.g., \texttt{Ctrl-,}, \texttt{Option-,}) helpful for navigating between significant code sections. One participant said, ``Dropping markers was a huge help. I could leave a marker at a critical point and come back to it without losing track," while another said they were ``constantly worrying about where [they] had left off" before they learned to start using the markers. The marker functionality was particularly useful for participants as they worked on code comprehension tasks with nested control structures. One participant noted that they were ``constantly using the markers, which made understanding if-else statements nested within loops much easier."



However, participants also noted an increase in cognitive load due to inconsistencies in feedback. When explaining their thought process out loud while working on a debugging task and focusing on determining which side of a array index square bracket their cursor was on, one participant explained, ``sometimes I expected a buzz to confirm my position in the code, but nothing happened, which was confusing." Another participant added, ``the audio feedback sometimes overlapped, especially when there were multiple errors, making it hard to focus on fixing one issue at a time." One participant muted the computer volume while receiving audio feedback, saying, ``there are times when the audio feedback is too much at once, I have to pause and figure out what is going on." Multiple participants also commented on the difficulty of remembering all of the keyboard shortcuts, noting ``the shortcuts are intuitive, but it took some time to remember all of them, it definitely added to my cognitive load initially" and ``I think there's a steep learning curve to get a hang of all the features which can be overwhelming for new users." Despite these challenges, the ability to toggle feedback granularity greatly helped. One participant noted, ``The option to switch between detailed and minimal feedback was great; it helped me focus on what I needed without getting overwhelmed." 

\begin{figure}[h] 
    \centering
	\includegraphics[trim={0.0cm 0.0cm 0.0cm 0.0cm},clip,width=0.8\textwidth]{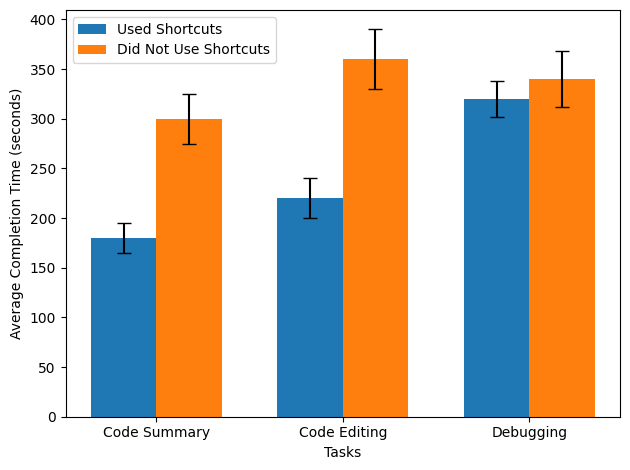}
	\vspace{-0.0em}
    \caption{Average Task Completion Time by Shortcut Usage.\label{fig:shortcutTiming}}
\end{figure}

\begin{figure}[h] 
    \centering
	\includegraphics[trim={0.0cm 0.0cm 0.0cm 0.0cm},clip,width=0.8\textwidth]{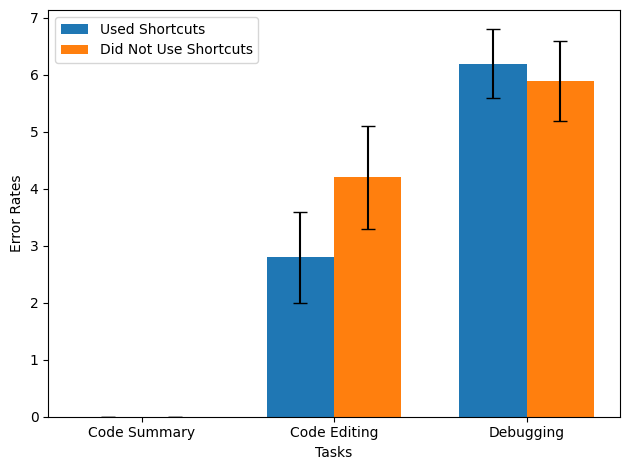}
	\vspace{-0.0em}
    \caption{Average Number of Errors Raised by Shortcut Usage.\label{fig:shortcutErrors}}
\end{figure}

While this study did not recruit enough participants and follow an appropriate experimental design to determine statistical significance for differences or establish causality, it is still meaningful to examine the differences in error rates and completion times between groups of participants who used HACI to approach the programming tasks in different ways. Figures \ref{fig:shortcutTiming} and \ref{fig:shortcutErrors} show the average completion time and error rate, respectively, for participants who used navigation shortcuts (i.e., jumping to either an absolute location or a marker) once or more in completing that task vs those who didn't use navigation shortcut functionality at all and navigated up and down using solely the arrow keys instead. We can see that participants who used shortcuts tended to make less errors and finish quicker when working on Code Summary and Editing tasks, while there is not as large of a difference for Debugging tasks. It is important to note, however, that those using the shortcuts may be more comfortable overall using HACI and may therefore have completed the tasks quicker regardless of their shortcut usage.

\section{Code Structure}
\label{sec:codeStructResults}

The haptic feedback corresponding to different code structures like indentations (ring and index finger buzzes) effectively aided in understanding nested structures and blocks of code. One participant mentioned, "Feeling the buzz when indentation changed helped me visualize the code structure better." Another participant added, "I appreciated the haptic cues for indentations, but sometimes they were too subtle or missed." Participants largely expressed the sentiment that the haptic feedback for indentations made it easier to follow nested structures in tasks where they were present.

However, the intensity and timing of the feedback sometimes lacked precision, causing moments of confusion. A participant explained this, ``The intensity of the buzzes needs to be adjustable. Sometimes it was too soft to notice." Another participant further explained, "When multiple haptic feedback cues happen close together, it was hard to distinguish them." This participant particularly noted this for moments like navigating onto a line with an error while navigating onto a bracket on that line, which would result in the haptic feedback for both occurring simultaneously. In response to facing a similar issue, one participant suggested, ``Different buzz patterns or intensities for various structures would be helpful."

Interestingly, participants discussed a lack of melding between the haptic and audio feedback when working to understand code structure, with one participant noting ``it feels separate from the diction, the haptic is for navigation first and then you read it out after." Another echoed this, noting, ``I used the haptics as a means of getting to where I wanted it to be read aloud to me." Additionally, while vertical line-to-line navigation was easily understandable via the haptic feedback, participants discussed a disconnect between the haptic and audio feedback as they sought to understand different parts of a given line horizontally. One participant reported that the ability to receive the haptic feedback about bracket position at moment the text-to-speech functionality reads a bracket out is critically lacking, for example. Instead, current functionality only triggered haptic feedback when the user moved the cursor over a character of interest themselves, regardless of what HACI was reading to them.


\section{Understandable Snapshots}

The functionality allowing for audible snapshots of the code, particularly the use of \texttt{Ctrl-G} to read the current line, was highly beneficial. Participants found it enabled quick verification of syntax and logic without manually reviewing every line, thus enhancing efficiency. One participant noted, ``The voice reading out the current line was very helpful, but it rushed through complex lines," highlighting both the benefit and a limitation of the feature.

However, the text-to-speech engine sometimes struggled with complex lines and variable names, necessitating multiple listens to understand fully. Reflecting on this, a participant stated, ``the text-to-speech engine struggled with complex variable names, making it hard to understand." This issue was compounded by pronunciation problems: ``the pronunciation of some terms was unclear, especially with complex variable names" and ``mispronounced terms made it difficult to follow along without having to repeat the line multiple times." These issues made it particularly difficult for participants to understand when the HACI was reading out a code snippet for indexing into an array (e.g., \texttt{arr[i]}). In response, a subject remarked, ``when you do operations on things, there are many better ways to make it into speech. For instance, where you index into the array, instead of `open square bracket i close square bracket' you could say something like `indexing into array at index i'."

\begin{figure}[h] 
    \centering
	\includegraphics[trim={0.0cm 0.0cm 0.0cm 0.0cm},clip,width=0.8\textwidth]{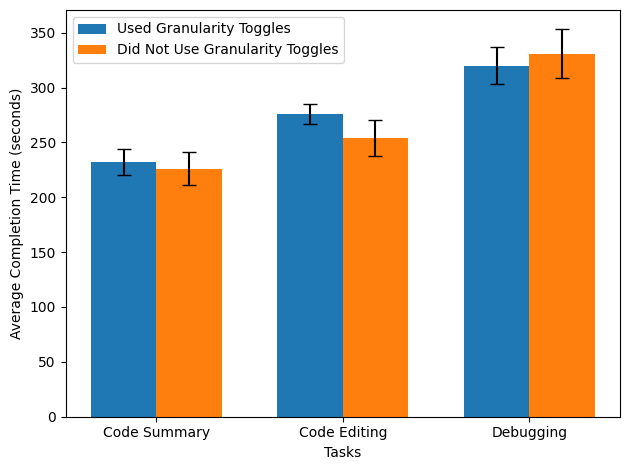}
	\vspace{-0.0em}
    \caption{Average Task Completion Time by Spoken Feedback Granularity Toggle Usage.\label{fig:granularityTiming}}
\end{figure}

\begin{figure}[h] 
    \centering
	\includegraphics[trim={0.0cm 0.0cm 0.0cm 0.0cm},clip,width=0.8\textwidth]{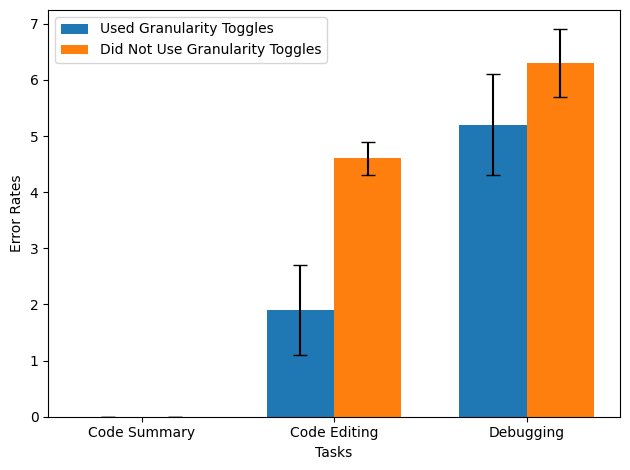}
	\vspace{-0.0em}
    \caption{Average Number of Errors Raised by Spoken Feedback Granularity Toggle Usage.\label{fig:granularityErrors}}
\end{figure}

The ability to toggle the granularity of audio feedback was crucial in managing information load. One participant appreciated this feature, saying, ``Being able to toggle the granularity of audio feedback was crucial. Sometimes I needed more detail, other times less." Another echoed this sentiment, noting, "The ability to adjust the verbosity of feedback helped manage the amount of information I was processing." Similar to the analysis provided in Chapter \ref{sec:cogLoadResults}, Figures \ref{fig:granularityTiming} and \ref{fig:granularityErrors} show the average completion time and error rate, respectively, for participants who toggled the granularity of their spoken feedback at least once vs. those who left it on the default setting. We can see that those who toggled granularity to adapt to the particular problem they were working on took slightly longer to complete their tasks on average, but encountered fewer errors. Once again, the caveat of potential confounding caused by the fact that those who toggled the granularity were likely those who were already more comfortable and more proficient at using HACI must be noted.

\section{Debugging}

Debugging with HACI showcased a mix of successes and challenges. The directional haptic feedback indicating where an error was located relative to the cursor (\texttt{Ctrl-E}) was innovative and well-received. One participant remarked, ``The haptic feedback indicating error location relative to the cursor was very innovative." This feature allowed for quicker error identification, as another participant observed, ``Knowing where the error was in relation to the cursor significantly sped up my debugging process."

Despite its benefits, the haptic feedback lacked specificity for different types of errors, even though HACI provides audio cues for error types, which some participants found limiting. One suggested, ``The system could use different patterns to indicate different types of errors." Additionally, there were concerns about the delay in error feedback. One participant expressed frustration, saying, ``The delay in error feedback sometimes made me miss the exact line, forcing me to backtrack." Furthermore, participants expressed a desire for directional error feedback to also indicate the horizontal position of the term causing the error on a given line. Participants frequently struggled locating where on a given line they need to navigate to in order to fix an error, and spent a considerable amount of time going through the line containing the error character-by-character until they hit a term which may match the error message they had heard.

Some participants primarily relied on audio feedback to debug instead of haptics, even to locate the line which an error was on. This was particularly achievable in debugging task 2 (presented in Appendix \ref{sec:interviewQuestions}, because the pre-determined misspelling of ``num" as ``nmu" in the task code was easily distinguishable from audio alone. Additionally, one participant approached debugging task 3 by noting the line on which he knew there was an error cause by an unrecognized variable name and having HACI read it out to him. He then noticed that the variable, which was ``message" misspelled as ``mesage," sounded the same both on the line where it was initially defined and the line where the error was. This enabled him to conclude that ``mesage" was missing an s, because any other spelling error would have cause HACI to read it out in a way that didn't sound like ``message."

Overall, while the haptic feedback for error detection was a step forward, further refinement is needed to enhance its effectiveness and usability.

\section{Cross-Cutting Results}

One of the major successes across the board was the integration of multi-modal feedback, which enriched the coding experience by diversifying the types of cues available. Participants found this particularly helpful in tasks involving structure navigation and error identification. One participant noted, ``The combination of audio and haptic feedback created a more immersive coding experience." Another stated, ``Having multiple types of feedback made it easier to understand the code environment."

After spending time using HACI, participants became more accustomed to working with the these feedback mechanisms, as one participant literally remarked, ``the integration of these feedback mechanisms feels natural over time." However, consistency and reliability of the feedback were crucial issues. One participant noted, ``There were instances where haptic feedback didn't trigger as expected, breaking my concentration." Ensuring consistent feedback is critical, as another participant emphasized, ``Ensuring consistent feedback is crucial, especially when the user relies on it for navigation and debugging."

Customization and usability were also highlighted as important factors. One participant suggested, ``Customizing feedback intensity and patterns based on user preference would enhance usability." Comfort and adjustability of the haptic gloves were also mentioned: ``The haptic gloves need to be comfortable and adjustable to accommodate different users, it's already a bit tight on my hand."

Overall, while HACI's multi-modal feedback provided a richer coding experience, addressing issues of consistency, reliability, and customization will be essential for its broader adoption and effectiveness.

\chapter{Discussion}
The pilot study provided valuable insights into the effectiveness and limitations of the Haptic-Audio Code Interface (HACI) in supporting visually impaired programmers. By analyzing both qualitative feedback from participants and quantitative data collected during the study, several key themes emerged that highlight the potential benefits and areas for improvement for HACI.

\section{Cognitive Load}

The use of HACI significantly reduced cognitive load by simplifying code navigation and syntax orientation. Participants appreciated the efficiency of shortcuts, such as \texttt{Option-1} for jumping to the top of the code, which helped them maintain control and reduce stress. The use of marker shortcuts was also beneficial for navigating between significant code sections. However, some participants noted an increase in cognitive load due to inconsistencies in feedback and the complexity of remembering all the keyboard shortcuts initially. These findings suggest that while HACI's navigation features can streamline the coding process, there is a need for improvements in feedback consistency and perhaps a more gradual introduction to the shortcuts to reduce initial cognitive load.

The quantitative data supported these qualitative insights, showing that participants who used shortcuts generally had lower error rates and faster completion times for Code Summary and Editing tasks. This suggests that effective use of HACI's navigational aids can significantly enhance coding efficiency, though the benefits were less pronounced for Debugging tasks. This indicates that while navigational aids are crucial, debugging requires additional support that addresses error specificity and real-time feedback.

\section{Code Structure}

HACI's haptic feedback for understanding code structure was well-received, helping participants visualize nested structures. However, there were calls for adjustable intensity and distinct patterns for different code structures to improve clarity. The integration of haptic and audio feedback was identified as an area needing improvement, with participants suggesting that the two feedback mechanisms should work more cohesively.

Participants felt that the haptic feedback often functioned separately from the audio feedback, which made it challenging to fully understand code structures. This indicates a need for more synchronized feedback mechanisms, where haptic cues are integrated with audio descriptions to provide a more cohesive understanding of the code.

\section{Understandable Snapshots}

The ability to receive audible snapshots of the code was a significant benefit, enabling quick verification of syntax and logic. Participants found the text-to-speech feature particularly useful, though there were issues with the engine struggling with complex lines and variable names. The option to toggle the granularity of audio feedback was crucial in managing information load, allowing participants to adjust the verbosity to their needs.

Quantitative data indicated that participants who adjusted feedback granularity encountered fewer errors, though they took slightly longer to complete tasks. This suggests that while more detailed feedback can slow down task completion, it enhances accuracy and reduces errors, highlighting the importance of customizable feedback settings in assistive coding tools.

\section{Debugging}

HACI's directional haptic feedback for error location was innovative and well-received, helping participants quickly identify error locations. However, the lack of specificity for different types of errors and occasional delays in feedback were noted as limitations. Participants suggested the need for distinct haptic patterns for various error types and more precise error location feedback.

Participants often relied on audio feedback to debug, indicating a preference for audio cues over haptic ones for error identification. This preference suggests that while haptic feedback is valuable, it should be complemented with more detailed audio feedback to enhance the debugging process.

\section{Cross-Cutting Results}

The integration of multi-modal feedback was a major success, providing a richer coding experience by diversifying the types of cues available. Participants found the combination of audio and haptic feedback helpful for structure navigation and error identification. However, the consistency and reliability of feedback were crucial issues that need addressing.

Customization and usability were also highlighted as important factors for broader adoption. Participants suggested that customizing feedback intensity and patterns based on user preferences would enhance usability. Additionally, the comfort and adjustability of the haptic gloves were noted as areas for improvement to accommodate different users.

Overall, while HACI's multi-modal feedback provided a richer coding experience, addressing issues of consistency, reliability, and customization will be essential for its broader adoption and effectiveness. Future research should focus on refining these aspects and exploring their impact on a more diverse population of visually impaired programmers. This study offers essential preliminary insights into the potential of haptic and audio feedback in supporting programming education for visually impaired users and lays the groundwork for further development and evaluation of HACI.

\section{Theoretical Implications}

The findings from this study contribute to the theoretical frameworks around multimodal learning and accessible computing education. By demonstrating that haptic and audio feedback can effectively convey information typically reliant on visual cues, this research supports the notion of sensory substitution and expansion in learning environments. It also challenges existing educational paradigms by proposing a more inclusive approach that considers the diverse sensory capabilities of learners.

\section{Practical Implications}

From a practical standpoint, the development and refinement of HACI provide a blueprint for the creation of more accessible coding environments. Educational institutions looking to accommodate a diverse student body, including those with visual impairments, can benefit from integrating such technologies into their curriculum. Moreover, this study outlines specific areas for improvement, such as feedback customization and error handling, which are critical for the user-friendly design of educational tools.

\chapter{Limitations}
The present study, aimed at evaluating the efficacy of a Haptic-Audio Code Interface (HACI) for programming education among visually impaired users, encounters several limitations that might influence the interpretation and generalization of the findings. One of the primary constraints is the participant demographic, as the study involved solely college students from the University of Chicago who had completed introductory Computer Science courses. Utilizing college students might not adequately represent the primary target audience—middle or high school students—who may have different cognitive abilities, learning styles, and familiarity with technology. The developmental and educational backgrounds of younger students could lead to different interactions with and benefits from the HACI, thus potentially limiting the generalizability of the study’s results to younger age groups.

Furthermore, all participants in the pilot study were sighted individuals temporarily blinded to simulate visual impairments. This simulation does not fully encapsulate the daily experiences of individuals who are permanently visually impaired, particularly those who have adapted long-term navigation and interaction strategies without sight. Sighted individuals may apply cognitive and spatial reasoning that does not authentically mimic those of individuals who have lifelong visual impairments. This discrepancy could affect the study's findings regarding the usability and effectiveness of the HACI, as sighted participants may not experience the same challenges and may not benefit from the interface in the same way as the visually impaired population would.

The recruitment process also posed significant limitations due to the very short timeline, which restricted the number of participants that could be included in the study. A condensed recruitment period can limit the diversity within the participant pool and affect the depth and breadth of data collected, potentially leading to less robust conclusions. Additionally, the rushed nature of participant recruitment might have impacted the availability and scheduling of participants for in-depth sessions, potentially affecting the consistency and reliability of the data collected.

The experimental setup, conducted in a controlled environment within the university premises, may not accurately reflect the diverse and sometimes less ideal conditions under which educational tools are utilized in real-world scenarios. Factors such as ambient noise, interruptions, and other real-life disturbances that could influence the functionality and user interaction with the HACI were not accounted for. This limitation could skew the perceived effectiveness of the interface when deployed in typical educational settings.

Moreover, the study might not fully explore how visually impaired individuals could uniquely benefit from the Haptic-Audio Code Interface (HACI) due to inherent differences in sensory processing capabilities between visually impaired and sighted individuals. Research indicates that visually impaired people often develop enhanced abilities in their other senses, particularly in auditory and tactile domains, which might influence their interaction with audio and haptic feedback systems \cite{collignon2009cross, sathian2010cross}. For example, studies have shown that visually impaired individuals can have better auditory memory and spatial awareness, which could allow them to manage simultaneous audio and haptic feedback more effectively than sighted individuals \cite{lessard1998early}. Additionally, they often exhibit a heightened ability to memorize and recall information after short exposure, likely a result of practice navigating environments through non-visual means \cite{roeder1999improved}.

These enhanced perceptual skills could mean that visually impaired users might not only manage but excel in utilizing the HACI's multimodal feedback for programming tasks, potentially leading to different or more rapid proficiency gains compared to sighted individuals simulating blindness. This aspect of sensory compensation suggests that the findings from the study, primarily involving temporarily blinded sighted participants, might under-represent the actual efficacy of the HACI when used by the visually impaired community. Future studies should consider these enhanced perceptual skills to tailor interface designs more closely to the real-world capabilities of visually impaired users, potentially exploring adaptive feedback systems that adjust based on user proficiency and sensory feedback preferences.

Despite these limitations, the study offers essential preliminary insights into the potential of haptic and audio feedback in supporting programming education for visually impaired users. To address the limitations, future research (as discussed in more detail in Chapter \ref{sec:futurework}) should strive to include a more representative sample of the target population, extend the recruitment and testing periods, test additional features of the HACI, and evaluate the system in various real-world educational environments to better understand its applicability and efficacy across different settings and among different user groups.

\chapter{Future Work}
\label{sec:futurework}
\section{HACI Improvements}

Future enhancements of the Haptic-Audio Code Interface (HACI) should focus on broadening its functionality and adaptability to better serve the diverse needs of visually impaired users in educational environments. One significant improvement could be the integration of multi-modal feedback. This would involve not only refining the existing haptic and audio feedback systems but also incorporating adjustable visual aids for users with partial vision or in settings where both visually impaired and sighted students are present. Features such as scalable text sizes, high contrast modes, and visual indicators can provide a more inclusive learning environment, making HACI a versatile tool in diverse classroom settings.

Another area ripe for development is the creation of customizable feedback profiles. Allowing users to tailor the types and intensities of feedback based on personal preferences or specific task requirements could enhance user experience and efficiency. Users could switch between profiles that, for instance, emphasize haptic feedback for code navigation and audio feedback for comprehension tasks, depending on their immediate needs or the specific challenges of the programming task at hand.

Advanced error handling and intelligent suggestions represent a further enhancement, transforming HACI from a passive tool into an active teaching assistant. By integrating a sophisticated AI-driven code analysis tool that not only detects errors but also suggests logical corrections, such as Github's Copilot \cite{Copilot}, HACI could significantly reduce the cognitive load on users and expedite the learning process. Such functionality would be particularly beneficial in helping users understand common programming pitfalls and learn to debug more effectively.

Expanding the range of programming languages supported by HACI is also essential. Currently focused on JavaScript, extending support to include popular languages such as Python, Java, and C++ would make HACI applicable to a wider curriculum and more accessible to students and educators in various programming disciplines. This expansion would not only increase HACI’s utility but also its adoption in computer science education globally.

Gesture-based controls could also be explored to enhance the interface’s accessibility. Implementing gesture recognition technology to allow users to execute commands or navigate through code via simple hand movements could provide an ergonomic and intuitive way to interact with the system, especially for users who may find continuous keyboard use cumbersome.

Additionally, developing peer collaboration features would allow visually impaired students to engage in group programming projects more seamlessly. Real-time collaboration tools integrated into HACI could foster an inclusive educational experience, enabling visually impaired students to participate fully alongside their sighted peers in interactive learning activities.

Lastly, incorporating cloud-based code management could revolutionize how users interact with HACI. By allowing users to save, retrieve, and manage their code projects via the cloud, HACI would support a flexible learning environment where students can continue their work across different devices and locations, facilitating both in-class learning and remote education.

Implementing these improvements would significantly enhance the functionality and appeal of HACI, making it a more comprehensive tool for programming education among visually impaired students and potentially increasing its adoption across educational institutions.

\section{Summer Camp User Study}

As part of future work, I propose the execution of a multi-week summer camp focused on using the Haptic-Audio Code Interface (HACI) with visually impaired students. This user study aims to explore the potential enhancements in accessibility and learning outcomes through extended use in a non-academic setting, while also allowing for the collection of detailed data on usability and effectiveness.

\subsection{Camp Design}
The summer camp will be designed to cater specifically to middle school students who are visually impaired. The curriculum will incorporate introductory programming concepts through project-based learning, using HACI to facilitate interaction and engagement. The camp will last for three weeks, allowing students to delve deeper into computer science topics and gain confidence in their programming skills.

\subsection{Curriculum and Activities}
The camp's curriculum will be structured around daily programming tasks, interactive lessons, and group projects that encourage collaboration and problem-solving. The activities will be designed to be engaging and fun, integrating game-like elements where possible to maintain a high level of interest and motivation among the students.

\subsection{Goals}
The primary goals of the summer camp will be to:
\begin{itemize}
    \item Assess the extended use of HACI in improving the coding skills of visually impaired students.
    \item Evaluate the accessibility features of HACI in a relaxed, educational setting outside of the typical classroom environment.
    \item Gather qualitative and quantitative data on how visually impaired students interact with coding environments and how such interactions can be improved through tailored tools like HACI.
\end{itemize}

\subsection{Assessment Methodology}
The effectiveness of the camp will be evaluated using a combination of observational studies, structured interviews, and direct assessments of student projects. This mixed-methods approach will allow for a comprehensive assessment of both the learning outcomes and the usability of HACI. Observations will focus on how students interact with the interface, noting any difficulties or barriers they encounter. Structured interviews will provide insights into the students' subjective experiences and perceptions of the programming process with HACI. Project assessments will evaluate the students' ability to apply their learned skills to complete specific tasks.

\subsection{Expected Outcomes}
The summer camp may provide valuable insights into the practical application of HACI in an educational setting. It is expected that students will show improvement in their understanding of programming concepts and their ability to navigate and manipulate code using audio-haptic feedback. The study may also highlight areas where HACI could be further improved to better serve the needs of visually impaired programmers.

\subsection{Future Extensions}
Based on the outcomes of the summer camp, further adaptations and iterations of HACI could be developed. Additional features could be integrated based on feedback from the camp participants, and further studies could be planned to follow up on the initial findings. This ongoing cycle of development and testing will help ensure that HACI evolves in a direction that truly enhances the educational experiences of visually impaired students in computing.

\chapter{Conclusion}
This thesis explored the development and evaluation of the Haptic-Audio Code Interface (HACI), a novel educational tool designed to improve accessibility and learning outcomes for visually impaired students in introductory programming courses. The integration of multimodal feedback—specifically haptic and audio cues—aimed to address several known barriers that visually impaired individuals face when learning to code. Through the implementation of HACI, this work sought to reduce cognitive load, enhance code structure comprehension, provide comprehensible auditory snapshots of code, assist in debugging through non-text feedback, and ensure user customizability to support a variety of learning styles and needs.

The pilot study presented in this thesis provides preliminary evidence supporting the effectiveness of HACI in enhancing the programming education experience for visually impaired students. Reported reductions in cognitive load were observed due to the intuitive design of navigation shortcuts and feedback mechanisms. Moreover, the unique use of haptic feedback in conjunction with auditory cues allowed for a deeper understanding of code structure and easier navigation through complex codebases without visual input.

Despite its successes, several challenges and limitations were identified. These include the need for more precise and reliable feedback mechanisms and the challenge of integrating such a system seamlessly into existing educational environments. Additionally, the learning curve associated with mastering new interfaces like HACI suggests that further refinement is needed to make these tools more accessible to novice users.

For future work, this thesis outlines several paths:
\begin{itemize}
    \item Enhancing the precision and reliability of feedback to reduce inconsistencies that disrupt the coding process.
    \item Expanding user studies to include a larger and more diverse group of visually impaired participants to further validate and refine the system.
    \item Developing comprehensive training modules to facilitate easier adoption and mastery of HACI functionalities.
    \item Exploring the integration of HACI with other programming languages and environments to broaden its applicability.
\end{itemize}

Ultimately, the Haptic-Audio Code Interface stands as a promising step toward more inclusive technology-driven education. By continuing to develop and refine such tools, we can ensure that visually impaired students are not only able to participate in computer science education but can also excel in this increasingly important field.

\section*{Acknowledgments}
This thesis would not have been possible without the guidance,
mentorship, and support of several people.  Before even starting my
first year at the University of Chicago, Professor Diana Franklin's
expertise and direct mentorship allowed me to learn about the
research process and begin my academic journey in computer science
education with the UChicago Computing for Anyone Lab.  As my
advisor, Diana has been crucial in helping me design and complete
this study from scratch while staying adaptable and informed.  I
also deeply appreciate Professor Marshini Chetty, for her
mentorship in providing me with more practical experience in
computer science education while I served as her Teaching Assistant,
and for making time to contribute her expertise as a reader on this
thesis.  I am deeply appreciative to my family members who have
supported me through the process of writing this thesis.  My mother
served as a fantastic source of inspiration and feedback
throughout, ensuring I was making consistent progress on this
project during a very stressful time, while providing advice from
her own experience completing her Master's thesis in Computer
Science twenty years ago.  It was my younger brother Krish's
undying intellectual curiosity that initially inspired me to take
on this project and pursue what I'm passionate about.
Additionally, I'm thankful to my friends and roommates for their
encouragement and flexibility as I dedicated such a large amount of
time to this project.  Finally, I am grateful to the student who
joined a programming class I was teaching five years ago,
determined not to let his visual impairment inhibit his love of
learning, for instilling in me the passion he had to defy barriers
put in the way of his intellectual journey and to ensure others can
do the same.  Thank you to you all.

\bibliographystyle{unsrt}
\bibliography{references}

\begin{thebibliography}{10}

\bibitem{organisation2016skills}
Organisation for Economic Co-operation and Development (OECD).
\newblock Skills for a digital world.
\newblock 2016.

\bibitem{schwieger2018reaching}
Dana Schwieger and Christine Ladwig.
\newblock Reaching and retaining the next generation: Adapting to the expectations of gen z in the classroom.
\newblock {\em Information Systems Education Journal}, 16(3):45, 2018.

\bibitem{bhalla2021digital}
Rohan Bhalla, Pinaz Tiwari, and Nimit Chowdhary.
\newblock Digital natives leading the world: paragons and values of generation z.
\newblock In {\em Generation Z Marketing and Management in Tourism and Hospitality: The Future of the Industry}, pages 3--23. Springer, 2021.

\bibitem{abesadze2020make}
Shorena Abesadze and David Nozadze.
\newblock Make 21st century education: The importance of teaching programming in schools.
\newblock {\em International Journal of Learning and Teaching}, 6(3):6, 2020.

\bibitem{vegas2020we}
Emiliana Vegas and Brian Fowler.
\newblock What do we know about the expansion of k-12 computer science education?
\newblock 2020.

\bibitem{zain2016integration}
Ismail~Md Zain, Balakrishnan Muniandy, and Wahid Hashim.
\newblock The integration of 21st-century learning framework in the asie instructional design model.
\newblock {\em Psychology Research}, 6(7):415--425, 2016.

\bibitem{alismail201521st}
Halah~Ahmed Alismail and Patrick McGuire.
\newblock 21st century standards and curriculum: Current research and practice.
\newblock {\em Journal of Education and Practice}, 6(6):150--154, 2015.

\bibitem{lamb2017key}
Stephen Lamb, Quentin Maire, and Esther Doecke.
\newblock Key skills for the 21st century: An evidence-based review.
\newblock 2017.

\bibitem{pires2020exploring}
Ana~Cristina Pires, Filipa Rocha, Antonio~Jos{\'e} de~Barros~Neto, Hugo Sim{\~a}o, Hugo Nicolau, and Tiago Guerreiro.
\newblock Exploring accessible programming with educators and visually impaired children.
\newblock In {\em Proceedings of the Interaction Design and Children Conference}, pages 148--160, 2020.

\bibitem{fields2014programming}
Deborah~A Fields, Michael Giang, and Yasmin Kafai.
\newblock Programming in the wild: trends in youth computational participation in the online scratch community.
\newblock In {\em Proceedings of the 9th workshop in primary and secondary computing education}, pages 2--11, 2014.

\bibitem{mountapmbeme2021teachers}
Aboubakar Mountapmbeme and Stephanie Ludi.
\newblock How teachers of the visually impaired compensate with the absence of accessible block-based languages.
\newblock In {\em Proceedings of the 23rd International ACM SIGACCESS Conference on Computers and Accessibility}, pages 1--10, 2021.

\bibitem{alotaibi2020teaching}
Hind Alotaibi, Hend S.~Al-Khalifa, and Duaa AlSaeed.
\newblock Teaching programming to students with vision impairment: Impact of tactile teaching strategies on student’s achievements and perceptions.
\newblock {\em Sustainability}, 12(13):5320, 2020.

\bibitem{stefik2017quorum}
Andreas Stefik and Richard Ladner.
\newblock The quorum programming language.
\newblock In {\em Proceedings of the 2017 ACM SIGCSE Technical Symposium on Computer Science Education}, pages 641--641, 2017.

\bibitem{Baker2015}
Catherine~M. Baker, Lauren~R. Milne, and Richard~E. Ladner.
\newblock Struct jumper: A tool to help blind programmers navigate and understand the structure of code.
\newblock volume 2015-April, pages 3043--3052. Association for Computing Machinery, 4 2015.

\bibitem{Albusays2016}
Khaled Albusays and Stephanie Ludi.
\newblock Eliciting programming challenges faced by developers with visual impairments: Exploratory study.
\newblock pages 82--85. Association for Computing Machinery, Inc, 5 2016.

\bibitem{Albusays2017}
Khaled Albusays, Stephanie Ludi, and Matt Huenerfauth.
\newblock Interviews and observation of blind software developers at work to understand code navigation challenges.
\newblock pages 91--100. Association for Computing Machinery, Inc, 10 2017.

\bibitem{mealin2012exploratory}
Sean Mealin and Emerson Murphy-Hill.
\newblock An exploratory study of blind software developers.
\newblock In {\em 2012 ieee symposium on visual languages and human-centric computing (vl/hcc)}, pages 71--74. IEEE, 2012.

\bibitem{huff2020examining}
Earl~W Huff, Kwajo Boateng, Makayla Moster, Paige Rodeghero, and Julian Brinkley.
\newblock Examining the work experience of programmers with visual impairments.
\newblock In {\em 2020 ieee international conference on software maintenance and evolution (icsme)}, pages 707--711. IEEE, 2020.

\bibitem{kane2014tracking}
Shaun~K Kane and Jeffrey~P Bigham.
\newblock Tracking@ stemxcomet: teaching programming to blind students via 3d printing, crisis management, and twitter.
\newblock In {\em Proceedings of the 45th ACM technical symposium on Computer science education}, pages 247--252, 2014.

\bibitem{Alotaibi2020}
Hind Alotaibi, Hend~S. Al-Khalifa, and Duaa AlSaeed.
\newblock Teaching programming to students with vision impairment: Impact of tactile teaching strategies on student's achievements and perceptions.
\newblock {\em Sustainability (Switzerland)}, 12, 7 2020.

\bibitem{smith2003nonvisual}
Ann~C Smith, Justin~S Cook, Joan~M Francioni, Asif Hossain, Mohd Anwar, and M~Fayezur Rahman.
\newblock Nonvisual tool for navigating hierarchical structures.
\newblock {\em ACM SIGACCESS Accessibility and Computing}, (77-78):133--139, 2003.

\bibitem{stefik2011design}
Andreas~M Stefik, Christopher Hundhausen, and Derrick Smith.
\newblock On the design of an educational infrastructure for the blind and visually impaired in computer science.
\newblock In {\em Proceedings of the 42nd ACM technical symposium on Computer science education}, pages 571--576, 2011.

\bibitem{Potluri2018}
Venkatesh Potluri, Priyan Vaithilingam, Suresh Iyengar, Y.~Vidhya, Manohar Swaminathan, and Gopal Srinivasa.
\newblock Codetalk: Improving programming environment accessibility for visually impaired developers.
\newblock volume 2018-April. Association for Computing Machinery, 4 2018.

\bibitem{Mountapmbeme2022}
Aboubakar Mountapmbeme, Obianuju Okafor, and Stephanie Ludi.
\newblock Addressing accessibility barriers in programming for people with visual impairments: A literature review, 3 2022.

\bibitem{roberts2011audio}
Dominic Roberts and Karlton Weaver.
\newblock Audio aids in source code.
\newblock {\em Retrieved September}, 19:2017, 2011.

\bibitem{armaly2018comparison}
Ameer Armaly, Paige Rodeghero, and Collin McMillan.
\newblock A comparison of program comprehension strategies by blind and sighted programmers.
\newblock In {\em Proceedings of the 40th International Conference on Software Engineering}, pages 788--788, 2018.

\bibitem{stefik2009sodbeans}
Andreas Stefik, Andrew Haywood, Shahzada Mansoor, Brock Dunda, and Daniel Garcia.
\newblock Sodbeans.
\newblock In {\em 2009 IEEE 17th International Conference on Program Comprehension}, pages 293--294. IEEE, 2009.

\bibitem{mountapmbeme2020investigating}
Aboubakar Mountapmbeme and Stephanie Ludi.
\newblock Investigating challenges faced by learners with visual impairments using block-based programming/hybrid environments.
\newblock In {\em Proceedings of the 22nd International ACM SIGACCESS Conference on Computers and Accessibility}, pages 1--4, 2020.

\bibitem{smith2000java}
Ann~C Smith, Joan~M Francioni, and Sam~D Matzek.
\newblock A java programming tool for students with visual disabilities.
\newblock In {\em Proceedings of the fourth international ACM conference on Assistive technologies}, pages 142--148, 2000.

\bibitem{hutchinson2018initial}
Joe Hutchinson and Oussama Metatla.
\newblock An initial investigation into non-visual code structure overview through speech, non-speech and spearcons.
\newblock In {\em Extended Abstracts of the 2018 CHI Conference on Human Factors in Computing Systems}, pages 1--6, 2018.

\bibitem{ludi2016exploration}
Stephanie Ludi, Jamie Simpson, and Wil Merchant.
\newblock Exploration of the use of auditory cues in code comprehension and navigation for individuals with visual impairments in a visual programming environment.
\newblock In {\em Proceedings of the 18th International ACM SIGACCESS Conference on Computers and Accessibility}, pages 279--280, 2016.

\bibitem{armaly2018audiohighlight}
Ameer Armaly, Paige Rodeghero, and Collin McMillan.
\newblock Audiohighlight: Code skimming for blind programmers.
\newblock In {\em 2018 IEEE International Conference on Software Maintenance and Evolution (ICSME)}, pages 206--216. IEEE, 2018.

\bibitem{stefik2007wad}
Andreas Stefik, Roger Alexander, Robert Patterson, and Jonathan Brown.
\newblock Wad: A feasibility study using the wicked audio debugger.
\newblock In {\em 15th IEEE International Conference on Program Comprehension (ICPC'07)}, pages 69--80. IEEE, 2007.

\bibitem{dorsey2014developing}
Rayshun Dorsey, Chung~Hyuk Park, and Ayanna Howard.
\newblock Developing the capabilities of blind and visually impaired youth to build and program robots.
\newblock In {\em 28th Annual International Technology and Persons with Disabilities Conference}, 2014.

\bibitem{capovilla2013teaching}
Dino Capovilla, Johannes Krugel, and Peter Hubwieser.
\newblock Teaching algorithmic thinking using haptic models for visually impaired students.
\newblock In {\em 2013 Learning and Teaching in Computing and Engineering}, pages 167--171. IEEE, 2013.

\bibitem{falase2019tactile}
Olutayo Falase, Alexa~F Siu, and Sean Follmer.
\newblock Tactile code skimmer: A tool to help blind programmers feel the structure of code.
\newblock In {\em Proceedings of the 21st International ACM SIGACCESS Conference on Computers and Accessibility}, pages 536--538, 2019.

\bibitem{React}
{React Javascript Library}.
\newblock \url{https://react.dev/}.
\newblock [Online; accessed 21-April-2024].

\bibitem{collignon2009cross}
Olivier Collignon, Patrice Voss, Maryse Lassonde, and Franco Lepore.
\newblock Cross-modal plasticity for the spatial processing of sounds in visually deprived subjects.
\newblock {\em Experimental brain research}, 192:343--358, 2009.

\bibitem{sathian2010cross}
K~Sathian and Randall Stilla.
\newblock Cross-modal plasticity of tactile perception in blindness.
\newblock {\em Restorative neurology and neuroscience}, 28(2):271--281, 2010.

\bibitem{lessard1998early}
Nadia Lessard, Michael Par{\'e}, Franco Lepore, and Maryse Lassonde.
\newblock Early-blind human subjects localize sound sources better than sighted subjects.
\newblock {\em Nature}, 395(6699):278--280, 1998.

\bibitem{roeder1999improved}
Brigitte Ro{\`E}der, Wolfgang Teder-Sa{\`E}leja{\`E}rvi, Anette Sterr, Frank Ro{\`E}sler, Steven~A Hillyard, and Helen~J Neville.
\newblock Improved auditory spatial tuning in blind humans.
\newblock {\em Nature}, 400(6740):162--166, 1999.

\bibitem{Copilot}
{GitHub Copilot}.
\newblock \url{https://github.com/features/copilot}.
\newblock [Online; accessed 21-April-2024].

\end{thebibliography}

\appendix

\chapter{Pilot Study Interview Questions and Guide}
\label{sec:interviewQuestions}
I now provide the interview guide used to conduct the pilot study interviews, including the guiding points for the interviewers, the specific interview questions, and the code content of each of the programming tasks.

\section{Introduction ($\sim 2$ minutes)}
\begin{itemize}
    \item Welcome the participant and briefly explain the purpose of the study.
    \item Assure the participant of confidentiality and the anonymization of their responses.
    \item Explain the structure of the interview and the tasks involved
\end{itemize}

\section{Fitting and Training ($\sim 5$ minutes)}

"How comfortable are you with the HACI gloves fitted to you?"

"Do the motors on each finger feel intrusive or natural during coding tasks?"

"Please share your initial thoughts on using haptic feedback for code navigation."

\section{Task 1: Summarize Existing Code ($7$ minutes)}

There will be a time limit of 7 minutes on this section, it is fine if subjects don’t complete all three functions. Provide each function one at a time, and provide the corresponding prompt:

\begin{itemize}
    \item \textbf{FunctionA}: "Examine the code in \texttt{functionA}. Describe its purpose and how it processes the input."
    \item \textbf{FunctionB}: "Review \texttt{functionB} and explain the logic it implements, particularly focusing on the conditional operation."
    \item \textbf{FunctionC}: "Analyze \texttt{functionC}. Provide insights into what this function calculates and the significance of the loop within it."
\end{itemize}

\begin{lstlisting}
function functionA(arr) { 
    let resultA = 0; 
    for (let i = 0; i < arr.length; i++) { 
        if (arr[i] % 2 === 0) { 
            resultA += arr[i]; 
        } 
    } 
    return resultA; 
} 

function functionB(x, y) { 
    if (x > y) { 
        return x - y; 
    } 
    else { 
        return y - x; 
    } 
} 

function functionC(n) {
    let resultC = 1; 
    for (let i = 1; i <= n; i++) { 
        resultC *= i; 
    } 
    return resultC; 
} 

let data = [1, 2, 3, 4, 5]; 
console.log(functionA(data)); 
console.log(functionB(10, 5)); 
console.log(functionC(5));
\end{lstlisting}

"How did the haptic feedback assist in understanding the structure and flow of this code?"

"Were there any parts of the code that felt more challenging to interpret with the HACI?"

\section{Task 2: Edit Existing Code and Write New Code ($\sim 5$ minutes)}

\textbf{Objective}: Write a new function named \texttt{modifyOutput} that takes a string as an argument and returns a modified version of the string by appending " - processed" at the end. Then, edit the \texttt{main} function to use \texttt{modifyOutput} and log the result to the console.

\textbf{Solution}:

\begin{lstlisting}
function modifyOutput(inputString) {
    // Your code here to modify the input string 
    Return inputString + " - processed"; 
} 

function main() { 
    let originalString = "Hello, World"; 
    // Use the modifyOutput function to modify the originalString 
    let modifiedString = modifyOutput(originalString); 
    // Log the modified string to the console
    console.log(modifiedString); 
} 

main();
\end{lstlisting}

"How did you integrate the new function into the Main function? Did the HACI provide clear feedback during this task?"

\section{Task 3: Debugging Multiple Small Functions ($\sim 10$ minutes)}

Each snippet contains a deliberate error that needs to be identified and corrected by the participants.

"For each snippet, please identify and correct the error. How does the haptic feedback guide you towards finding these errors?"

Snippet 1: Index Out of Bounds Error

\begin{lstlisting}
function accessElement() {
    const elements = [1, 2, 3, 4, 5];
    console.log(elements[5]);  // Error: Array index out of bounds
}

accessElement();
\end{lstlisting}

Snippet 2: Misspelled Variable

\begin{lstlisting}
function calculateSum() {
    const numOne = 10;
    const numTwo = 20;
    const result = numOne + nmuTwo;  // Error: 'nmuTwo' is misspelled 
    console.log(result);
}

calculateSum();
\end{lstlisting}

Snippet 3: Something Not Declared

\begin{lstlisting}
function printMessage() {
    const message = "Hello, World!";
    console.log(mesage);  // Error: 'mesage' is not declared; it's a typo
}

printMessage();
\end{lstlisting}

"Which type of error was easiest to identify with the HACI, and why?"

"Did the HACI's feedback on errors align with your expectations based on the training session?"

\section{Reflective Questions ($\sim 5$ minutes)}

"Reflecting on all tasks, what were the biggest challenges you faced while using the HACI?"
"How does the HACI's method of debugging compare to your usual debugging practices? Are there aspects you found more efficient or more challenging?"

"Can you describe any particular moments during the tasks where the HACI significantly aided or hindered your understanding of the code?"

"Considering your experience with traditional screen readers and the HACI, how do you perceive the potential of haptic feedback in programming education and practice for visually impaired individuals?"

\end{document}